\documentclass[10pt,journal,compsoc]{IEEEtran}
\usepackage{balance}

\usepackage[utf8]{inputenc}
\usepackage{graphicx}
\title{Answering Uncertain, Under-Specified API Queries Assisted by Knowledge-Aware Human-AI Dialogue}
\author{Qing~Huang,
        Zishuai~Li,
        Zhenchang~Xing,
        Zhengkang~Zuo,
        Xin~Peng,
        Xiwei~Xu,
        Qinghua~Lu
\IEEEcompsocitemizethanks{\IEEEcompsocthanksitem Q. Huang, Z. Li, Z. Zuo are with School of Computer Information Engineering, Jiangxi Normal University, China.\protect
\IEEEcompsocthanksitem Q. Huang and Z. Li are co-first authors, Z. Zuo is the corresponding author(zuo803@jxnu.edu.cn).
\IEEEcompsocthanksitem Z. Xing, X. Xu and Q. Lu are with the CSIRO's Data61, Australia. 
\IEEEcompsocthanksitem X. Peng is with the School of Computer Science and Shanghai Key Laboratory of Data Science, Fudan University.
}
}
\date{October 2022}
\usepackage{graphicx}
\usepackage{float}
\usepackage{array}
\usepackage{tabularx}
\usepackage{multirow}
\usepackage{amsmath}
\usepackage{amssymb}
\usepackage{threeparttable}
\usepackage{setspace}
\usepackage{ulem}
\usepackage{color}
\usepackage{colortbl}
\usepackage{xcolor}
\usepackage{soul}
\usepackage{booktabs}
\usepackage{picins}

\usepackage{algorithm} 
\usepackage{algpseudocode}
\usepackage{hyperref}
\usepackage{caption}

\definecolor{mygray}{gray}{.9}
\definecolor{mypink}{rgb}{.99,.91,.95}
\definecolor{mycyan}{cmyk}{.3,0,0,0}
\definecolor{myyellow}{RGB}{255,230,204}
\definecolor{mybule}{RGB}{218,232,252}
\definecolor{mygreen}{RGB}{213,232,212}
\definecolor{titleColor}{RGB}{102,102,102}

\begin{document}
\IEEEtitleabstractindextext{%
\begin{abstract}
Developers' API needs should be more pragmatic, such as seeking suggestive, explainable, and extensible APIs rather than the so-called best result.
Existing API search research cannot meet these pragmatic needs because they are solely concerned with query-API relevance.
This necessitates a focus on enhancing the entire query process, from query definition to query refinement through intent clarification to query results promoting divergent thinking about results.
This paper designs a novel Knowledge-Aware Human-AI Dialog agent (KAHAID) which guides the developer to clarify the uncertain, under-specified query through multi-round question answering and recommends APIs for the clarified query with relevance explanation and extended suggestions (e.g., alternative, collaborating or opposite-function APIs).
We systematically evaluate KAHAID. 
In terms of human-AI dialogue efficiency, it achieves a high diversity of question options and the ability to guide developers to find APIs using fewer dialogue rounds.
For API recommendation, KAHAID achieves an MRR and MAP of 0.769 and 0.794, outperforming state-of-the-art methods BIKER and CLEAR by at least 47\% in MRR and 226.7\% in MAP. For knowledge extension, KAHAID obtains an MRR and MAP of 0.815 and 0.864, surpassing ZaCQ by at least 42\% in MRR and 45.2\% in MAP.
Furthermore, we conduct a user study.
It shows that explainable API recommendations, as implemented by KAHAID, can help developers identify the best API approach more easily or confidently, improving inspiration of clarification question options by at least 20.83\% and the extensibility of extended APIs by at least 12.5\%.
\end{abstract}

\begin{IEEEkeywords}
Developers' API Need, Knowledge Graph,  Human-AI Dialogue, API Recommendation, Multi-Round Question Answering.
\end{IEEEkeywords}}

\maketitle
\IEEEdisplaynontitleabstractindextext
\IEEEpeerreviewmaketitle

\IEEEraisesectionheading{\section{Introduction}}
\IEEEPARstart{D}{evelopers}' API (short for Application Programming Interface) needs are no longer just a matter of finding so-called best API for certain programming tasks. 
To minimize API misuse, they must consider several aspects such as the API's specific usage context, relations to cooperative APIs, and confusing APIs with subtle differences. \cite{Ren2020DemystifyOA}.
As a result, it is desirable that API search methods should guide developers to clarify the vague question intent, provide diverse and suggestive APIs for different needs, interpret the search results, and extend other potentially useful API knowledge~\cite{Eberhart2021DialogueMF}.
This expectation reveals some pragmatic API needs, which seek suggestive, explainable, and extensible API recommendation and knowledge discovery rather than simply presenting so-called best APIs.
Meeting these pragmatic API needs not only helps developers select the ideal APIs for their needs, but it also inspires and broadens their thinking, such as exploring alternative or better solutions, discovering previously unknown API knowledge.

Existing API search methods, however, are incapable of meeting the above-mentioned pragmatic API needs.
Given an API query, they measure query-API relevance based on keyword matching~\cite{Haiduc2011OnTE, Thung2013AutomaticRO, Rahman2016RACKAA} or embedding similarity~\cite{Mikolov2013DistributedRO,Ye2016FromWE}, and return the top-k related APIs as a list of ``best'' answers.  
These answers are only useful if the API query can clearly describe the actual query intent, but even then, they fall far short of the pragmatic API needs because the answers are viewed as independent of each other with little or no explanation.
Furthermore, an API query in the real world is highly likely uncertain and under-specified.
In order to clarify the query intent, the query can be extended based on word co-occurrence~\cite{Huang2018APIMR}.
However, without any human intervention, the extended words are very likely irrelevant and may potentially introduce noise that blur the query intent~\cite{Gu2016DeepAL, Xie2020APIMR}.
It follows that API search should shift from its current narrow focus on query-API relevance to enhancing the entire search process, from query definition to query refinement through intent clarification, to query results promoting divergent thinking about the results.

This query process is vividly reflected in the social-technical information seeking on online forums.
For example, on Stack Overflow, a question-answering (Q\&A) thread usually includes a number of comments that help clarify uncertain and under-specified API questions and discuss technical details and trade-offs~\cite{Zhang2021ReadingAO,Ren2019DiscoveringEA}, in which the developer often gains more than just a single API, but becomes familiar with various APIs' specific usage contexts and their cooperation and differences \cite{Liu2021APIRelatedDI}.
Although social-technical information seeking aligns well with the pragmatic API needs and supports much better question clarification, intent understanding and answer explanation through developer interactions and knowledge sharing, the question-answering process relies on human interaction and engagement, which cannot immediately respond to the developers' needs.

In this work, we design a novel Knowledge-Aware Human-AI dialogue agent (KAHAID) as the first step towards integrating the immediate response capability of API research and the interaction, clarification, explanation, and extensibility capability of social-technical information seeking.
Given an uncertain, under-specified query, KAHAID interacts with the developer to clarify the query intent through multi-round question answering and returns APIs with relevance explanation and extended knowledge for the clarified query.

Each dialogue round clarifies a certain aspect of the query by asking a question with a list of options.
To automatically generate meaningful clarification questions with diverse options, we construct an API behavior knowledge graph (KG) that extracts API actions, objects, constraints and diverse functional and semantic relations from API documentation.
To support efficient dialogue process, we design an information-gain based decision tree algorithm over the underlying knowledge graph to prioritize the questions posted to the developer and minimize dialogue rounds. 
Finally, KAHAID identifies APIs in the knowledge graph based on the developer's answers to its clarification questions, and present not only the most relevant API but also extended suggestions (e.g., alternative, cooperating or opposite-function APIs).
All recommended APIs come with the explanation of how they are related to the query and other APIs.
This result explanation can enhance the developer's trust in the recommended APIs and allow them to make informed choice among APIs.

We have implemented KAHAID for Java SDK APIs.

To evaluate the effectiveness of human-AI dialogue and the quality of API recommendations by KAHAID, we have developed 
three test sets for testing.
The first is the dataset we reused from BIKER’s manually created test dataset, the second is randomly selected 1k
sample SO posts to alleviate potential human bias.
To simulate questions and answers in real situations, we added a third test dataset, which contains manually selected 60 questions with ground-truth APIs from SO.
In terms of dialogue efficiency, KAHAID achieves high semantically diversity of question options (the average diversity between any two options is 74.9\%) and the ability to guide developers using fewer dialogue rounds to find APIs (the average number of rounds required for a query to find an answer is no more than three).
Our evaluation confirms that KAHAID has a strong ability for API recommendation and knowledge extension.
For API recommendation, KAHAID achieves a mean reciprocal rank (MRR) and mean average precision (MAP) of 0.769 and 0.794 respectively, and this outperforms the two state-of-the-art API search approaches BIKER and CLEAR by at least 47\% in MRR and 226.7\% in MAP.
In terms of knowledge extension, KAHAID achieves an MRR and MAP of 0.815 and 0.864 respectively, which surpasses the state-of-the-art conversation-based code search method, ZaCQ, by at least 42\% in MRR and 45.2\% in MAP.

Furthermore, we conducted a user study in which 12 Java developers were divided into two groups using different tools to find APIs for 8 programming tasks (derived from the selected 60 SO questions) using KAHAID and ZaCQ, respectively.
Using KAHAID improves the inspiration of clarification question options by at least 20.83\% and the extensibility of extended APIs by at least 12.5\%.
These results also confirm that explainable API recommendation can assist developers in selecting the best API method more easily and confidently. 

Overall, this paper makes the following contributions:

1) Conceptually, we are aware of pragmatic API needs, which include suggestive, explainable, and extensible API recommendation and knowledge discovery.
We believe that API search should shift from only finding the best APIs to enhancing the entire query process by providing a variety of potentially useful and enlightening knowledge.
Driven by this conception, we design KAHAID to achieve an interactive, illuminating, explainable, and extensible exploratory discovery.

2) We create an API behavior knowledge graph that includes API actions, objects, constraints, functional relations and semantic relations to generate clarification questions with multiple options. 
These questions can prompt developers to discover better options or help them determine their actual needs.

3) Over the underlying knowledge graph, we design an information-gain based decision tree algorithm to prioritize the questions posted to the developer, minimize question answering rounds, and guide the developer gradually clarify the vague question intent.


4) We evaluate KAHAID's ability to meet pragmatic API needs in terms of variety, guidance, extensibility, and interpretability.
Our data package can be found here \footnote{\href{https://github.com/Codeshuai/Answering-uncertain-under-specified-API-query-assisted-by-knowledge-aware-human-AI-dialogue}{https://github.com/Answering-uncertain-under-specified-API-query-assisted-by-knowledge-aware-human-AI-dialogue}}.

\vspace{-3mm}
\section{MOTIVATING EXAMPLE}
\label{Section:2}
\subsection{API Search Status}
In order to find the APIs for developers' needs, a common solution is to use the natural language description of the API need as a query, and use API search approaches to obtain some candidate APIs whose documentation is similar to the query.
During the search process, developers may seek diverse, explainable, guided, and extensible API recommendations, rather than just a single API.

Consider a \href{https://stackoverflow.com/questions/4871051/how-to-get-the-current-working-directory-in-java}{Stack Overflow question} where a developer needs a Java API to get the current working directory.
Let us consider how to satisfy this API need using API search over API documentation (assume that the answers like this SO post do not exist). 
If the search query is specific, the API search will be able to find the most relevant API, but it likely loses extensibility and miss other potentially useful APIs.
For example, assume the developer issues a specific query ``get absolute path string of current working directory in Java'',  BIKER~\cite{Huang2018APIMR} (a query expansion API search tool) can find the most relevant API directly (i.e., \textit{java.io.File.getAbsolutePath}) by matching the query with the API description. 
However, its results does not include \textit{java.nio.file.Path.toAbsolutePath}, a Java new IO API which may also satisfy the need ``get the current work directory''. 
But the description ``Returns a Path object representing the absolute path'' of \textit{java.nio.file.Path.toAbsolutePath} is not similar to the very specific query, so it will not be returned.
As a result, the developer may miss the opportunity to use a new API based on the API search results.

\begin{figure*}[h]
    \centering
    \includegraphics[width=1\textwidth]{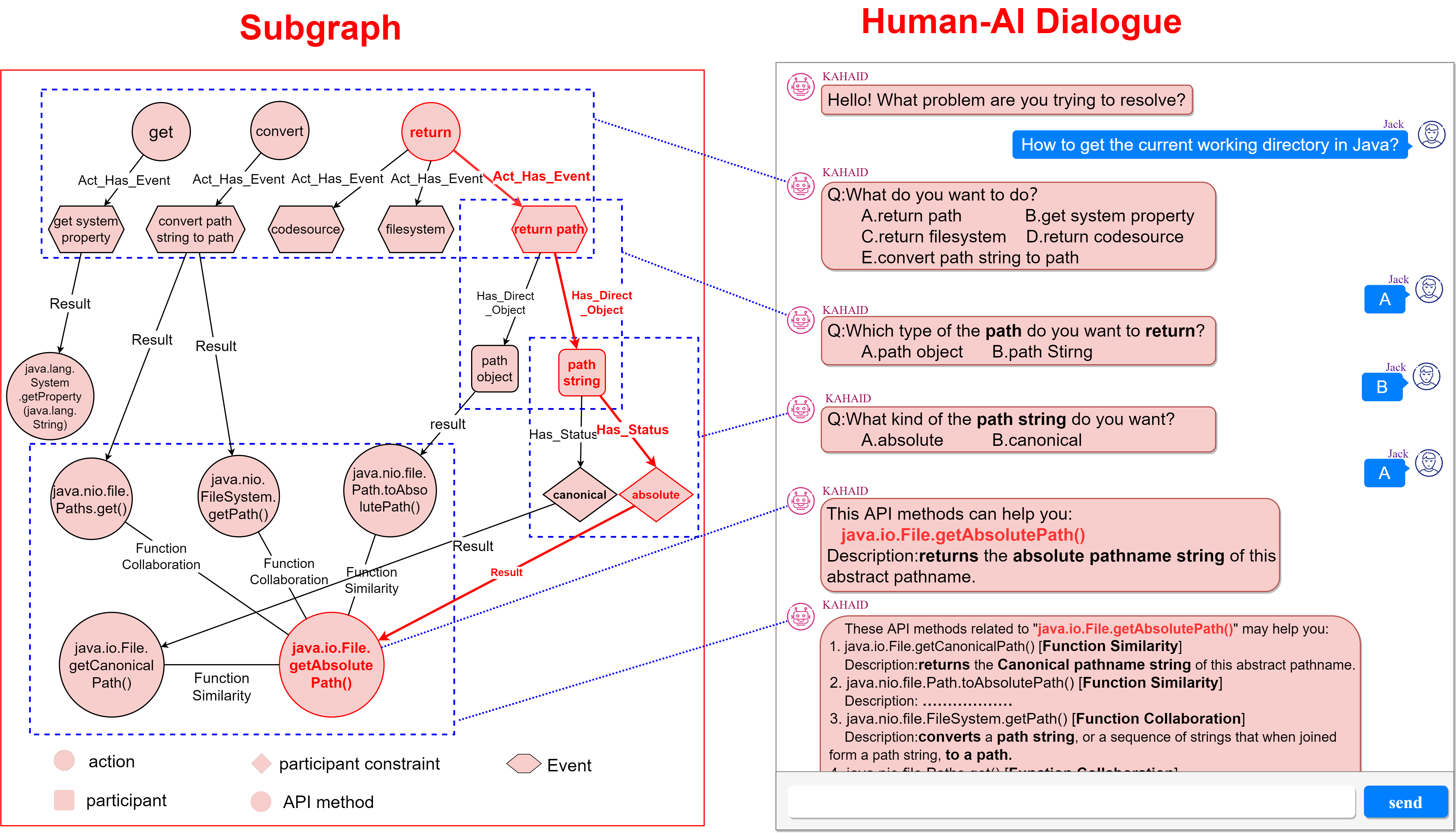}
    \footnotesize
     Note: The red text and lines in the Subgraph indicate the critical path necessary for obtaining the best API.
    \caption{An Example of API Recommendation with the help of Query-Related API Behavior Subgraph and Human-AI Dialogue}
    \label{Fig:motivating example}
\end{figure*}

On the other hand, if the search query is too broad, the API search tool very likely cannot find the relevant APIs.
For example, using the SO question title ``how to get the current working directory in Java'' as a query, BIKER returns a chaotic set of APIs.
In top 10, only one API \textit{java.io.File.getCanonicalPath} (returns the canonical path string of the current directory) is relevant but this API is ranked seventh.
The poor search results actually reveal another issue of API search, i.e., lack of interpretability, making it difficult for the developers to interpret the search results and determine their relevance and trustworthiness.
For example, the top-1 ranking API is \textit{java.io.File.list}, but it is difficult to understand how \textit{java.io.File.list} could be used to get the current working directory, as it only lists all files in the current working directory.
It is also difficult to understand why this irrelevant API is ranked first while the relevant API \textit{java.io.File.getCanonicalPath} is ranked seventh.

\vspace{-3mm}
\subsection{API Knowledge Seeking on Stack Overflow}
Different from API search, question-answering process on Stack Overflow offers a completely different experience.
The Stack Overflow question ``How to get the current working directory in Java\footnote{\href{https://stackoverflow.com/questions/4871051/how-to-get-the-current-working-directory-in-java}{\textcolor{blue}{https://stackoverflow.com/questions/4871051/}}}'' received 70 comments, many of which help to clarify the vague question intent. 
For example, one comment asks ``What is it you're trying to accomplish by accessing the working directory? Could it be done by using the class path instead?''.
This comment clarifies the action is ``accomplish'', the object of this action is ``what'', and whether or not the constraint on this event is ``by using the class path''.
Another comment says ``Knowledge of the current working directory is important for all relative paths. If you think that is irrelevant make sure you always access files via some absolute path.''
This comment clarifies whether the constraint on the object ``path'' is ``all relative'' or ``some absolute''.

The question receives diverse answers, including directly relevant APIs such as \textit{java.io.File.getAbsolutePath} and \textit{java.io.File.getCanonicalPath}, as well as extended APIs like \textit{java.nio.File.toAbsolutePath} with similar functionality, and \textit{java.nio.File.Paths.get} as a cooperative API that converts the path string from \textit{java.io.File.getAbsolutePath} to a path object.
Meanwhile, these answers and comments on Stack Overflow also provide explanation of the recommended APIs, drawn from the API documentation.
As a result, these explanations reinforce the developer's understanding and trust of the recommended APIs in the answers.
However, as the Q\&A process relies on human inputs, those diverse answers were provided across 4 years.

\subsection{API Search Assisted by Human-AI Dialogue}
\label{Subsection:B}

In this work, we aim to assist API search with human-AI dialogue which simulates the capability of intent clarification and result explanation and extension as the Q\&A process on Stack Overflow, and meanwhile can provide the immediate response to the search query.
The Q\&A process on Stack Overflow is underpinned by human knowledge of API behaviors and relations.
In the same vein, API search with human-AI dialogue needs to be supported by a knowledge graph of API behaviors which represents API actions, objects, constraints and various API functional and semantic relations in a graph like the example shown in Fig.~\ref{Fig:motivating example}.
Given a search query, the agent interacts with the developer to clarify the query intent until it finds some APIs.
It presents the found APIs with the explanation how they are related to the clarified query and each other.

Fig.~\ref{Fig:motivating example} illustrates an example of this knowledge-graph supported human-AI dialogue process for API search.
The developer Jack initially asks an under-specified question ``How to get the current working directory in Java'', for which the current API search tool (e.g., BIKER~\cite{Huang2018APIMR}) returns poor results.
However, the agent KAHAID, based on the API behavior knowledge graph, determines it needs some clarification of actions first, because several different actions are somewhat related to ``get working directory''. 
It asks ``what do you want to do?'' with a list of options extracted from the knowledge graph, such as ``return path'', ``return filesystem'', ``return codesource'' or ``convert path string to path''.
Jack replies ``return path''.
With the clarified action, KAHAID determines it needs some further clarification about the type of path to be returned, either ``path object'' or ``path string'', again based on the knowledge graph.
Jack replies ``path string'', which triggers another round of clarification ``which kind of path string'', either absolute or canonical.
Jack replies ``absolute''.
Through three rounds of human-AI dialogue, the intent of the initial under-specified question becomes clear, which leads KAHAID to an API \textit{java.io.File.getAbsolutionPath}.

KAHAID goes beyond \textit{java.io.File.getAbsolutionPath} to identify other relevant APIs using semantic relations in the knowledge graph. Specifically, KAHAID locates two functionally similar APIs, namely \textit{java.io.File.getCanonicalPath} and \textit{java.nio.file.Path.toAbsolutePath}, as well as two functionally cooperative APIs, namely \textit{java.nio.file.Paths.get} and \textit{java.nio.file.FileSystem.getPath}.
Instead of simply presenting some APIs, KAHAID provides a concise explanation for each recommended API.
First, it shows the API's functionality description and highlights the keywords that it uses as clarification options.
This helps Jack determine why these APIs are recommended and how they are relevant to his question.
Second, KAHAID shows the relations of the extended APIs and the most relevant API, such as function similarity and function cooperation.

To sum up, API search assisted by knowledge-aware human-AI dialogue will create an exploratory search process that is suggestive, explainable, and extensible.
It can lead to effective and serendipitous API recommendation and knowledge discovery.

\vspace{-0.318cm}
\section{APPROACH}
\begin{figure*}[t]
    \centering
    \includegraphics[width=0.9\textwidth]{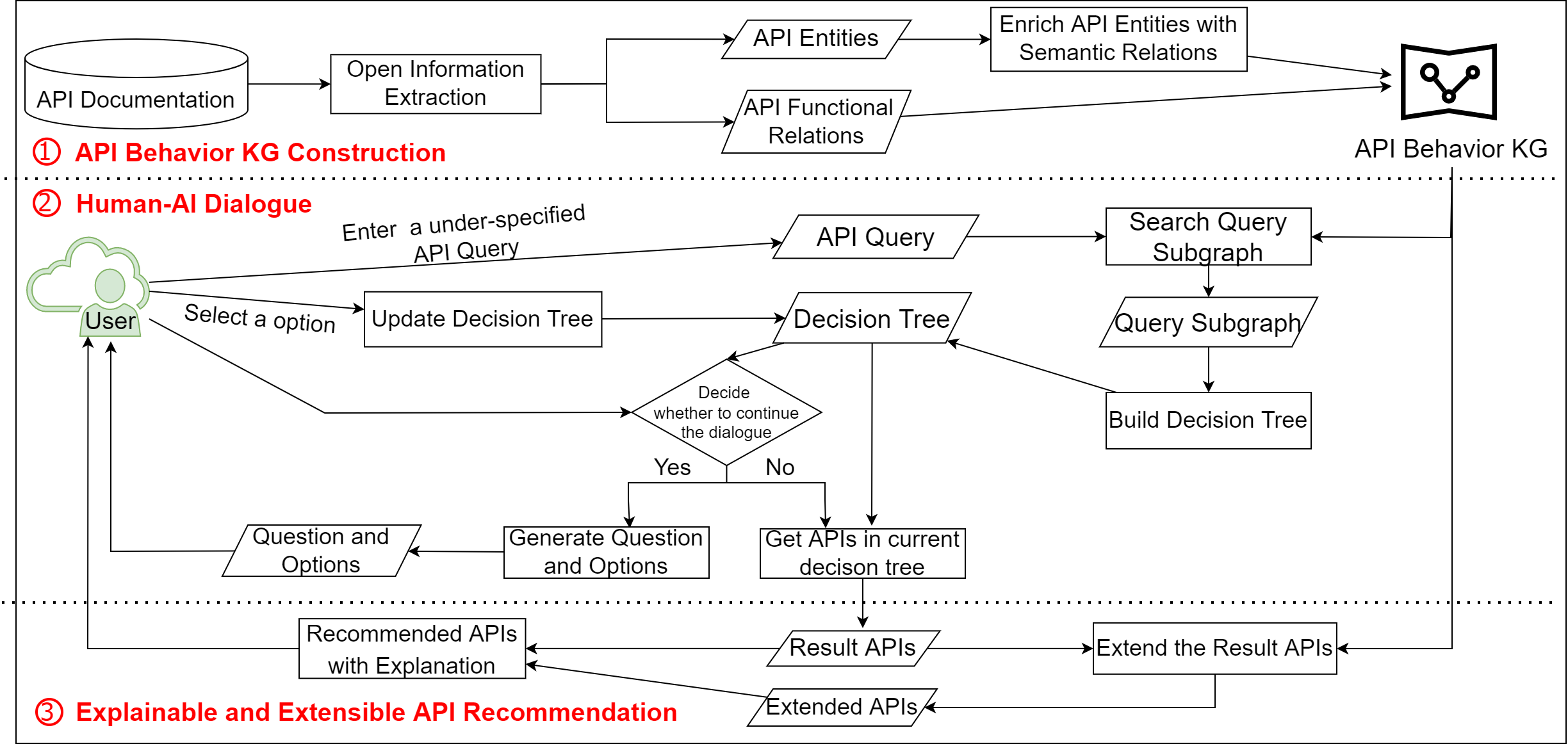}
    \caption{Overall Framework of KAHAID}
    \label{Fig:overview}
\end{figure*}

Fig.~\ref{Fig:overview} depicts the three stages of KAHAID design: API Behavior KG Construction, Human-AI Dialogue, and Explainable and Extensible API Recommendation.

In the first stage, we extract API entities (such as API actions, objects, events and constraints) and diverse functional relations from API documents using natural language processing (NLP) techniques, and then enrich the various semantic relations of API entities. 
Finally, the API behavior KG is constructed.

In the second stage, given an under-specified API query, KAHAID uses the subgraph search method to retrieve the Query-related API behavior subgraph containing various functional relations from KG. 
Based on the subgraph, KAHAID builds a decision tree using the information-gain based decision tree algorithm, and initiates a dialogue based on this tree.
If the user don`t decide to continue the dialogue, KAHAID generates the result APIs directly and proceeds to the third stage; otherwise, KAHAID generates a clarification question with a variety of options to assist the user in clarifying the uncertain aspects of the query. After the user selects an option that represents a specific branch of the decision tree, KAHAID updates the decision tree based on the user's choice by merging and pruning the decision tree.

In the third stage, KAHAID returns the explainable result API and any extended APIs that have API semantic relations with the result API.

\subsection{API Behavior KG Design and Construction}
As stated in Section \ref{Subsection:B}, API search with Human-AI dialog need to be supported by a knowledge graph of API behaviors which represents API actions, objects, constraints and various API functional and semantic relations in a graph.
The following section describes how to design and build the API behavior knowledge graph.

\subsubsection{API Knowledge Source}

The API documentation contains a wealth of knowledge about API behavior, typically documented in API descriptions.
These descriptions are frequently used to answer questions on Stack Overflow.
For example, a question titled ``How to get the path of a running JAR file?''\footnote{\href{https://stackoverflow.com/questions/320542/}{\textcolor{blue}{https://stackoverflow.com/questions/320542/}}} is posted.
The sixth respondent endorsed API \textit{``java.nio.file.Paths.get()''} by referencing the functional event mentioned in the API description\footnote{\href{https://docs.oracle.com/javase/8/docs/api/java/nio/file/Paths.html\#get-java.lang.String-java.lang.String...-}{\textcolor{blue}{https://docs.oracle.com/javase/8/docs/api/java/nio/file/Paths-.html\#get-java.lang.String-java.lang.String...}}}, thereby supporting the recommendation.

In this work, we download the JDK 1.8 API reference documentation \footnote{\href{https://docs.oracle.com/javase/8/docs/api}{\textcolor{blue}{https://docs.oracle.com/javase/8/docs/api}}} and parse each API class's HTML file to extract API methods and their descriptions following the method mentioned in \cite{Liu2020GeneratingCB}, resulting in API method-description pairs.
Finally, we build a Java API dictionary with 30,200 API method-description pairs.

\vspace{-2mm}
\subsubsection{KG Design}
Our KG design is driven by two types of knowledge: clarification question related knowledge and result extension related knowledge.

\noindent\textbf{Clarification question related knowledge}: our KG includes the API behavior knowledge necessary for clarifying the following three types of questions in the under-specified queries.

The first type of question is event-oriented and serves to clarify the user's goals.
For example, given a question ``What do you want to do?'', we can derive the \textbf{Event} entity (``convert a path string to a path''), and further divide the Event entity into the \textbf{Action} entity (``convert'') and the \textbf{Object} entity (``path string'' or ``path'') because the event is composed of the action and the object.
As a result, the \uline{Act Has Event} relation naturally arises between the entities Action and Event, while the \uline{Has Direct Object} and \uline{Has Preposition Object} relations naturally arises between the entities Event and the Object. For example, ``path string'' and ``path'' are the direct and prepositional objects of ``convert''.

The second type of question is object constraint-oriented and serves to clarify the type or statue of object. For example, given a question ``What type of the path string do you convert?'', we can derive the \textbf{Object Constraint} entity, despite the fact that there is no modifier for ``path string'' in this real API description.
The Object Constraint relation will naturally arise between the Object and the Object Constraint entities.
However, this relation can be refined into two relations based on object modifiers.
The definitions and examples for two relationships are listed, with the type of relation underlined and the Object Constraint entity italicized.
The one is the \uline{Has Status} relation when the object modifiers are adjectives (ADJ), verbs (VERB), quantifiers (NUM) or adverbs (ADV).
For example, ``\textit{absolute} path string'' and ``\textit{canonical} path string''.
The another is the \uline{Has Type} relation when the object modifiers are noun (NOUN) or proper noun (PROPN), such as java built-in data types {[}``byte'',``int/integer'', ``float'', ``char'', ``boolean'', ``double'', ``long'', ``short''{]}.
For example, ``writes a \textit{double} value, which is comprised of four bytes, to the output stream.''

The third type of question is event constraint-oriented and serves to clarify the type or statue of event. 
For example, given a question ``What is the condition under which a path string is converted to a path?'' from this question, we can derive the \textbf{Event Constraint} entity (``when joined form a path string''). To be converted into ``a path'', ``the sequence of strings'' must meet this constraint.
The Event Constraint relation naturally arises between the Event entity and the Event Constraint entity.
And this relation can be refined into nine types based on the semantic roles (such as Locatives, Directional, Manner, Extent, Temporal, Goal, Purpose Clauses, Secondary Predication, Adverbials) proposed by PropBank \cite{Bonial2010PropBankAG}.
Each relation's definitions and examples are listed, with the underlined part representing the type of relation and the italicized part representing the Event Constraint entity.
\vspace{-2mm}
\begin{itemize}
    \item \uline{Has Location relation} (Location of event occurrence), for example, ``finds all the keys of the streams \textit{in this applet context}.''
    \item \uline{Has Direction relation} (Direction of event occurrence), for example, ``moves the focus \textit{down one focus traversal cycle}.''
    \item \uline{Has Manner relation} (Manner of event executed), for example, ``adds the specified component to the layout, \textit{using the specified constraint object}.''
    \item \uline{Has Extent relation} (Scope of event), for example, ``\textit{fully} parses the text producing a temporal object.''
    \item \uline{Has Temporal relation} (Temporal of event occurrence), for example, ``converts a path string, or a sequence of strings that \textit{when joined form a path string}, to a path.''
    \item \uline{Has Goal relation} (Target object of event), for example, ``fetches the command list \textit{for the editor}.''
    \item \uline{Has Purpose relation} (The effect event can achieve), for example, ``destroys the orb \textit{so that its resources can be reclaimed}.''
    \item \uline{Has Result relation} (The form of the event outcome), for example, ``gets a representation of the current choice \textit{as a string}.''
    \item \uline{Has Condition relation} (Condition of event occurrence), for example, ``returns the window object representing the full-screen window \textit{if the device is in full-screen mode}.''
\end{itemize}

As for the other semantic roles in PropBank, we do not consider any of them for three reasons.
First, not all those semantic roles are Event constraints, for example, the use of Adjectival (ADJ) is limited to noun annotations.
Second, Some semantic roles have no practical meaning, such as Discourse (DIS), which is a token (such as "however," "to," "as well as,") that connects one sentence with the previous sentence.
Third, some semantic roles, such as Cause Clauses (CAU), which are used to explain causes, are not part of the API's behavior.


Furthermore, the API entity is derived because all of the entities mentioned above are related to the API \textit{``java.nio.file.Paths.get()''}.
As a result, the \uline{API Has Event} relation will naturally emerge between the API and the Event entities.

So far, we've deduced API entities, event relations, event constraint relations, and object constraint relations.
And we summarize these three types of relations as the functional relation (also known as intra-relation), which describes the API's functionality from fourteen aspects.

\textbf{Result extension related knowledge:}
In addition to the API behavior knowledge, our KG also needs the API semantic knowledge necessary for clarifying the relationship between the result API and other APIs.
As a result, we employ seven types of API semantic relations proposed by Yuan et al.\cite{Huang2022112PK}, namely, \uline{Function Similarity, Function Opposite, Function Replace, Function Collaboration, Logic Constraint, Behavior Difference, Efficiency Comparison}.
Here we only consider these seven method-level API semantic relations and ignore the other non-method-level API semantic relations mentioned in Yuan's work\cite{Huang2022112PK} because the Human-AI dialogue we propose is focused on API methods.



\subsubsection{KG Construction}
\label{Sub:3.1.2}
Following KG design, we extract API entities and relations required for KG from API descriptions based on semantic and syntactic roles.

\textbf{Step 1: Semantic and Syntactic Role Annotation.}
Given a specific API and its description, we annotate the sentence with semantic roles using the natural language tool AllenNLP\cite{Gardner2018AllenNLPAD}, and obtain the functional parts with functional semantic roles and the constraint parts with constraint semantic roles.
The functional parts are then assembled into a functional statement in the correct order, and it is part-of-speech tagged to yield grammatical parts with six different syntactic roles, such as verb, direct object, preposition, preposition object, direct object's modifier, and preposition object's modifier.

\textbf{Step 2: Entity and Functional Relation Extraction based on Annotation.}
We organize these grammatical parts and constraint parts into six entities and fourteen functional relations based on the following rules:

(1) The API entity is the qualified name of a API method that corresponds to the API description.

(2) The Event entity is made up of grammatical parts with the syntactic roles ``verb + direct object'' or ``verb + preposition + preposition object''.

(3) The Action entity is the grammatical part with the syntactic role ``verb''.

(4) The Object entity is the grammatical part with the syntactic role ``direct object'' or ``preposition object''.

(5) Along with the formation of four entities (API, Event, Action, Object), the four event relations (API Has Event, Act Has Event, Has Direct Object, Has Preposition Object) are formed naturally.

(6) The Object Constraint entity is the grammatical part with the syntactic role ``direct object's modifier'' or ``preposition object's modifier''; its object modifier determines the type of object constraint relation which it forms with the Object entity.
If the object's modifier is an adjective (ADJ), verb (VERB), quantifier (NUM) or adverb (ADV), the object constraint relation is the Has Status relation; otherwise, it is the Has Type relation.

(7) The Event Constraint entity is the constraint part with constraint semantic roles; its semantic role determines the type of event constraint relation which it forms with the Event entity.
One constraint semantic role corresponds to one event constraint relation, that is, ARGM-LOC corresponds to Has Location, ARGM-DIR to Has Direction, ARGM-MNR to Has Manner, ARGM-EXT to Has Extent, ARGM-TMP to Has Temporal, ARGM-GOL to Has Goal, ARGM-PRP to Has Purpose, ARGM-PRD to Has Result, and ARGM-ADV to Has Condition.

\begin{figure*}[h]
    \centering
    \includegraphics[width=1\textwidth]{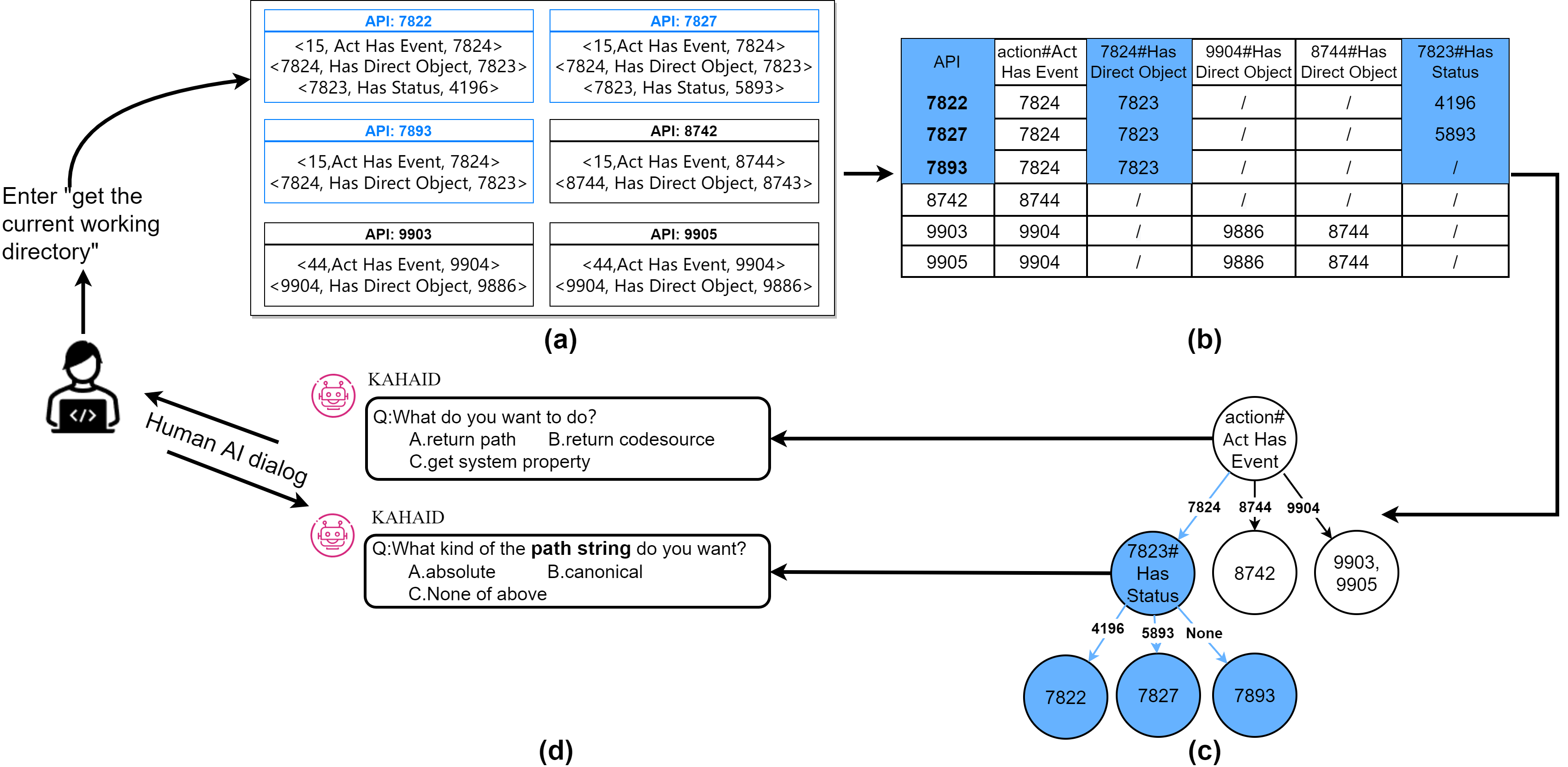}
    \footnotesize
     Note: Figure (a), (b), and (c) illustrate the step-by-step generation of clarifying dialogues in Figure (d) through KAHAID, with emphasis on the blue partition.
    \caption{Clarification Question Generation Process}
    \label{Fig:decision tree}
    \vspace{-5.0mm}
\end{figure*}

So far, six types of entities and fourteen kinds of functional relations (known as intra-relation) have been extracted from an API and its description.
To help accurately retrieving functional relations related to a specific API in subsequent sections, we have set up a FUNCTION property for each API entity and stored the extracted functional relations from the API description in this property, as shown in Fig.\ref{Fig:decision tree}-a.

\textbf{Step 3: Semantic Relation Extraction.}
As for the seven types of API semantic relations (known as inter-relation), we refer to the API-Task knowledge graph proposed by Yuan et al.\cite{Huang2022112PK}, which consists of many API relation triples, each of which is of the form $ \langle $API name, semantic relation Name, API name$ \rangle $.
If the API names of two API method entities match two API names in a triple, this triple's semantic relation becomes the relation between these two entities.
Despite the fact that the API name in our API entity is a full qualified name (FQN) while the API name in the triple of API-Task KG is a simple name, we can successfully match because the simple name can be converted to the FQN based on the three points listed below.

(1) the method-level APIs to match all have parentheses, a key distinguishing symbol, for example, \textit{``update()''}.

(2) For the simple method name made up of two or more tokens, one token represents a method and the other represents a class.
The package can be reasoned out from the method and the class, and then we combine method, class, and package to get FQN.
For example, we can infer the ``java.io'' package from the ``InputStream'' class and the ``read()'' method of \textit{``InputStream.read()''}.

(3) Even if two simple method names have only one token, they belong to the same class if they appear together.
As a result, we can reason class, then package from the class and the method, and finally FQN.
For example, for the triple $ \langle $\textit{update()}, Function Similarity, \textit{doFinal()}$ \rangle $, since two methods ``update'' and ``doFinal'' appear together, we can infer two classes ``\textit{javax.crypto.Cipher}'' and ``\textit{java.crypto.Mac}'', and further two triples with FQNs, namely, $ \langle $\textit{javax.crypto.Cipher.update()}, Function Similarity, \textit{javax.crypto.Cipher.doFinal()}$ \rangle $ and $ \langle $\textit{java.crypto.Mac.update()}, Function Similarity, \textit{java.crypto.Mac.doFinal()}$ \rangle $.

\subsection{Human-AI Dialogue Process}
The Human-AI dialogue process is underpinned by Subgraph Search and Clarification Question Generation, which makes KAHAID to generate clarification questions alongside diverse options in each round of dialogue, assisting users in clarifying their under-specified API query intent.

\vspace{-3mm}
\subsubsection{Subgraph Search}
\label{sub:3.1}
KAHAID, with the help of the API behavior KG, can generate a subgraph for each query. 
This subgraph contains extensive API function knowledge, allowing multiple rounds of human-AI dialogue from fourteen aspects (technically, fourteen different types of functional relations) to clarify ambiguous queries and recommend appropriate APIs.

Given an under-specified API query, KAHAID searches for the top-N candidate APIs using established API search models (such as DeepAPI \cite{Gu2016DeepAL}, BIKER \cite{Huang2018APIMR}, and RreMA \cite{Xie2020APIMR}).
The API search model is set up on the same data source as our KG (Java SE 8 API document) for keyword matching, enabling alignment of candidate APIs with at least N API entities in our KG.
It should be noted that a single candidate API may correspond to multiple API entities in our KG, as candidate API names lack parameters while API entities in our KG include them. 
For example, the candidate API ``\textit{java.nio.file.paths.get}'' can match multiple APIs with different arguments in our KG, such as ``\textit{java.nio.file.paths.get(java.net.URI)}'' and ``\textit{java.nio.file.paths.get(java.lang.string, java.lang.string)}''.

KAHAID extends the matched APIs to gather more query-related API entities by fetching API Has Event relations from the FUNCTION property of matched API entities and searching our KG for entities with these relations.

Finally, KAHAID considers the retrieved API entities and their functional relations in FUNCTION properties as a query-related API behavior subgraph, similar to the example in Fig.~\ref{Fig:motivating example} (left).
The functional relations are represented as $\langle$e1, r, e2$\rangle$, indicating a relation called r between API entities e1 and e2. 
Note that the \textit{API Has Event} relation is excluded of the subgraph as it has been used to extend the API (as mentioned above). 

\vspace{-2mm}
\subsubsection{Clarification Question Generation}
\label{Sub:3.2}
In this process, we use the subgraph built-in \ref{sub:3.1} to generate clarification questions and a variety of options for each round of the dialogue process.
For each functional relation $\langle$e1, r, e2$\rangle$ in the subgraph, e1 and r form a specific aspect (a aspect that under-specified query needs to clarify), and e2 is the option of that aspect.
This means that we can generate clarification questions and options by selecting the best aspect from the subgraph.

To generate clarification questions (CQs) from the subgraph, we refer two types of CQ generation approaches: the task extraction approach\cite{Eberhart2022GeneratingCQ} and the decision tree-based approach\cite{CastleGreen2020DecisionTA}.
The task extraction approach identifies grammatical roles from the comment and selects the most frequent ones as aspects to clarify.
The templates are then created based on the aspects to generate the CQs.
However, this approach lacks guidance as it does not consider aspect priorities.
In contrast, the decision tree-based approach can classify function aspects with uncertain options by estimating probabilities of candidate APIs\cite{Hssina2014ACS}, providing better guidance.
In our current work, we employ the decision tree-based approach. 
This approach comprises four steps, with the first three involving the construction of a complete decision tree, and the fourth step entailing the generation of CQs based on the decision tree.

\textbf{Step 1: Organize each API and functional relation into an attribute table.}
Constructing an attribute table is a pivotal step in the construction of a decision tree, where APIs and functional relations are organized in a two-dimensional format. 
The attribute table facilitates convenient comparison of different aspects, aiding in the identification of the most critical aspect that requires clarification in a query.
We refer to API entities and their functional relations as API\{$\langle$e1, r, e2$\rangle$,...\} (as shown in \ref{Fig:decision tree}-a), which are then converted to the attribute table with one API column and multiple aspect columns. 
This process is depicted in the progression from Fig.~\ref{Fig:decision tree}-a to b.

As shown in Fig.~\ref{Fig:decision tree}-b, the first column is API column, which contains all API number in Fig.\ref{Fig:decision tree}-a.
Columns after first column are aspect columns.
For a functional relation $\langle$e1, r, e2$\rangle$ in Fig.\ref{Fig:decision tree}-a, we use its aspect “e1\#r” as the column name and option ``e2'' as the column value.
Note that if r in $\langle$e1, r, e2$\rangle$ is an \textit{Action Has Event} relation, we use “action\#Has Event” directly as the column name (rather than “e1\#r”).
For example, in Fig.\ref{Fig:decision tree}-a, the APIs enclosed in blue boxes all have the same functional relationship $\langle$15, Has Event, 7824$\rangle$. 
We use "action\#Has Event" (instead of "15\#Has Event") as the second column name in Fig.\ref{Fig:decision tree}-b.

\textbf{Step 2: Select critical aspect based on the attribute table and split the subgraph to get sub-datasets.}
When construct a decision tree, it is necessary to the critical aspect(the first line of aspect column) from the attribute table. 
A critical aspect assists in minimizing the height of a decision tree, resulting in improved efficiency in the question-answering process.
Two decision tree strategies, ID3 \cite{Quinlan2004InductionOD} and C4.5\cite{Quinlan1992C45PF}, are available for selecting aspect.
C4.5 employs information gain ratio for aspect dialogue round selection, but increases average dialogue rounds compared to ID3, which may not expedite user intention clarification.
In our current implementation, we use ID3 to generate the CQ by selecting the aspect column with the highest information gain. 
Higher information gain corresponds to more options or APIs associated with the question, which helps reduce dialogue rounds and guide users to clarify their intentions quickly.

The information gain of the aspect column is calculated by Eq.\ref{equ.gain}, which is the difference between the information entropy of all candidate APIs and the information entropy of the current aspect column.
Entropy measures the uncertainty of a random variable and characterizes the impurity of examples.
Higher entropy indicates more information content \cite{Quinlan2004InductionOD}.

The information entropy of all candidate APIs is calculated by Eq.\ref{equ.I} where m is the number of candidate APIs and  $ P_{i} $ is the likelihood that the i-th API appears in all candidate APIs.
The information entropy of current aspect column is calculated by Eq.\ref{equ.E}, where k is the type of column values in the aspect column, ``$ API_{1j},...,API_{mj} $'' represents m APIs associated with the j-th column value.

\vspace{-1mm}
\begin{equation}
    Gain(aspect)=I(API_{1},...,API_{m})-E(aspect)
    \label{equ.gain}
\end{equation}

\vspace{-1mm}
\begin{equation}
    I(API_{1},...,API_{m})=-\sum_{i=1}^{m}P_{i} \log_{2}{P_{i}}
    \label{equ.I}
\end{equation}

\vspace{-1mm}
\begin{equation}
    E(aspect) = \sum_{j = 1}^{k}[\frac{|API_{1j},...,API_{mj}|}{|API|}\times I(API_{1j},...,API_{mj})]
    \label{equ.E}
\end{equation}

Based on the calculation, we identify the aspect column with the highest information gain from the attribute table. 
We refer the aspect of this column as the current node in the decision tree, with the different column values serving as edges connecting to current node (including null values).

In order to generate child nodes for each edge, it`s need to partition the subgraph. 
We group API\{$\langle$e1, r, e2$\rangle$,...\} with the same column value together to form sub-datasets.
For example, the three API\{$\langle$e1, r, e2$\rangle$,...\} enclosed in the blue box in Graph A form a sub-dataset, which grouped together based on the column value 7824.
At this point, we have completed the construction of one layer of nodes and edges in the decision tree.


\textbf{Step 3: Recursively repeat Step 1 and Step 2 to construct a complete decision tree that supports dialogue.}
In order to construct a complete decision tree, for each sub-dataset, we repeat the Step 1 and Step 2 recursively to build child nodes until the following stopping criteria are met.
1) The sub-dataset only contains one API.
2) The sub-dataset contains multiple APIs with identical functionalities.
When the stopping criteria are met, we refer all APIs in current sub-dataset as the current node, which is set as a leaf node in the decision tree.
It's worth noting that when building a new attribute table, we remove the aspects that have already been selected to prevent generating redundant clarification questions.

\textbf{Step 4: After building the decision tree, KAHAID generates the human-AI dialogue process.}
Specifically, we generate the CQ and diversity options by using the decision tree's current node and all of its edges.
Based on different aspect of the current node, CQ can be generated in the following templates: 

\vspace{-2mm}
\begin{itemize}
    \item \textit{action\#Act Has Event}: What do you want to do?
    \item \textit{object\#Has Status}: What kind of the \{object\} do you want?
    \item \textit{object\#Has Type}: Which type of the \{object\} do you want?
    \item \textit{event\#Has Location}: Where the \{event\} will be done?
    \item \textit{event\#Has Direction}: Where is the direction of \{event\}?
    \item \textit{event\#Has Manner}: How would you prefer to \{event\}?
    \item \textit{event\#Has Extent}: How far would you want to \{event\}?
    \item \textit{event\#Has Temporal}: When do you have to \{event\}?
    \item \textit{event\#Has Goal}: Which object do you want to serve by \{event\}?
    \item \textit{event\#Has Purpose}: Which purpose do you want to satisfy by \{event\}?
    \item \textit{event\#Has Result}: What is the form of the results of \{event\}?
    \item \textit{event\#Has Constraint}: Under which condition can \{event\}?
\end{itemize}

Following the user's selection of options, the decision tree is pruned, leaving only the user-selected branch's sub-tree.
If the sub-tree has only one node, the answer will be recommended directly.
Otherwise, a new round of CQ and options will be generated.
During the dialogue process, users can manually stop generating new round of CQ and options at any point.
In such cases, we consider all APIs included in the sub-tree as the recommend API sequence.

Fig.~\ref{Fig:decision tree}-c to d illustrates how CQs are generated for the query "get the current working directory?" based on the decision tree.
KAHAID generates two rounds of dialogue to clarify events and objects separately.
Two round of CQ are generated by two node (action\#Act Has Even, 7823\#Has Status), and options for CQ are generated by node`s edge.


\subsection{Result Extension and Interpretation}
\label{sub:3.3}
After multiple dialogue rounds, KAHAID would further extend the result APIs and interpret all result APIs.
Specifically, KAHAID locates the corresponding API entity in the sub-graph for each result API and searches for the semantic relations (e.g., \textit{Function Similarity}, \textit{Function Collaboration}) that contain this API from KG.
The extended API is then the API corresponding to another entity in these semantic relations.
For example, in Fig.~\ref{Fig:motivating example} (left), the result API \textit{java.io.File.getAbsolutePath} exists in a semantic relation $\langle$java.io.File.getAbsolutePath, Function Similarity, java.io.File.getCanonicalPath$\rangle$, where \textit{java.io.File.getCanonicalPath} serves as the extended API.
Note that duplicate APIs, whether among the extended APIs or with the result APIs, are removed.
These extended APIs enable users to be inspired to find new ways to solve problems and meet their API needs more comprehensively.

Meanwhile, KAHAID provides API descriptions and highlights keywords to make the result APIs more interpretable. 
There are two types of highlighted keywords. 
The first type includes semantic relation names between result APIs and extended APIs, which explain how the result APIs are related to each other. 
These keywords appear after the extended API. 
For example, "Function Similarity" appears after the extended API \textit{java.nio.file.Path.toAbsolutePath}, indicating that it implements functionality similar to API \textit{java.io.File.getAbsolutePat}. 
The second type of highlighted keywords come from functional aspects with various options that were clarified during the dialogue process. 
KAHAID searches the decision tree for the path that includes the API in question and other nodes and edges in between. It then obtains keywords from this path and highlights them in the API description. 
For example, if the query is "How to get the current working directory in Java?" and the API \textit{java.io.File.getAbsolutePat} is included in the path, KAHAID obtains keywords such as "returns", "absolute", and "path string" from this path and highlights them in the API description. 
This result explanation helps increase user trust in the result APIs and enables informed API selections.

\section{EVALUATION DESIGN}
\subsection{Research Questions}
To evaluate the performance of KAHAID, we conduct a series of experiments to answer the following research questions:

\textbf{RQ1}: How Performance is KAHAID compare with Existing API Search Approaches~\cite{Huang2018APIMR, Wei2022CLEARCL}

\textbf{RQ2}: How Performance is KAHAID compare with Existing Human-AI Dialogue Approach~\cite{Wei2022CLEARCL}

\textbf{RQ3}: How semantically diverse are the question options generated during the Human-AI Dialogue?

\textbf{RQ4}: What factors can improve the dialogue efficiency?

\subsection{API Behavior Knowledge Graph Building}
We collect data from JAVA API documentation to construct an API Behavior KG in order to evaluate KAHAID.
First, we download the JDK 1.8 API reference documentation \footnote{\href{https://docs.oracle.com/javase/8/docs/api}{\textcolor{blue}{https://docs.oracle.com/javase/8/docs/api}}} and parse each API class's web page to extract API methods and their descriptions, yielding 30,200 API method-description pairs for building a Java API dictionary.

Second, following the knowledge graph construction method described in Section 3.1, we extract 48,420 entities (which includes 25,020 API entities and 23,400 other entities such as Action, Event, Object, Object Constraint and Event Constraint) and 89,160 relations from the 30,200 API method-description pairs, and save 25,020 API descriptions.
Note that if an event entity cannot be extracted from a description for each API method-description pair, we do not treat this API method as an API entity and do not save this description or any other entities extracted from this description.
For a more in-depth discussion, see Section 6's threat analysis.

Finally, we group these extracted entities and relations into an API Behavior KG, which was stored in Neo4j\footnote{\href{https://neo4j.com/}{\textcolor{blue}{https://neo4j.com/}}}.

\vspace{-2mm}
\subsection{Dataset}
To comprehensively evaluate the performance of KAHAID, we obtained two types of data from Stack Overflow (SO) to gather experimental queries and ground-truth APIs.
The first type of data consists of queries paired with a single ground-truth API for each query, allowing us to assess KAHAID's ability to recommend the most appropriate API. 
The second type of data comprises queries paired with multiple ground-truth APIs, including the best one and related ones, which enables us to evaluate KAHAID's effectiveness in knowledge expansion.

\subsubsection{Dataset with Single ground-truth API}
For dataset with single ground-truth API, we reuse SO data from the state-of-the-art approach BIKER~\cite{Huang2018APIMR}, which were collected from the official data dump of SO by following criteria: 1) the question is related to Java JDK programming, 2) the question should have a positive score, and 3) at least 1 answer to the question contains API entities and the answer’s score should be positive.

The APIs were extracted from the code snippets in markdown scripts of the accepted answers in SO. 
In a markdown script, code snippets are wrapped by $\langle$code$\rangle$ tags. One can use regular expressions to localize the code snippets and further extract the APIs.
BIKER also provided a test dataset for evaluating its performance. 
The test data contains 413 questions along with their ground-truth APIs. 
We use the title of these 413 questions as the query for evaluation.
Since the ground-truth APIs only have a single API, we refer it as a best API.

Note that, BIKER’s test dataset mainly contains SO posts with high quality, which cannot reflect the overall quality of SO posts. 
Thus, we have also created a random test datasets which contain randomly selected SO posts for removing human bias (the same dataset creat manner has also been used in the comparison of CLEAR and BIKER in CLEAR’s paper~\cite{Wei2022CLEARCL}).

\subsubsection{Dataset with Multiple ground-truth API}
\label{query and APIs}
For dataset with Multiple ground-truth API, we manually collect 60 API-related questions with multiple ground-truth APIs from 60 SO posts following BIKER`s criteria~\cite{Huang2018APIMR}.

The questions posted on Stack Overflow typically receive answers and comments.
To assess KAHAID's knowledge expansion ability, ground-truth APIs must include the best API and extended APIs with seven semantic relations to best API.
These relations, Function Similarity, Function Opposite, Function Replace, Function Collaboration, Logic Constraint, Behavior Difference, and Efficiency Comparison, are described in Section \ref{Sub:3.1.2}. 
Unlike BIKER, we consider APIs from all positively-scored answers and comments, not only from accepted answers.
This is because the accepted answer often only includes a single API, while non-accepted answers and comments often provide more extended APIs.

To obtain the best API and extended APIs, we asked an expert and two external annotators (two master students familiar with Java but unaffiliated with this work) to review all answers and comments in specific question posts. 
After a 30-minute training on classifying non-best API methods, the two annotators reviewed the API methods separately. 
The expert made the final decision if there was disagreement. 
Finally, we obtained 60 queries and 298 ground-truth API methods, including 60 best API methods and 238 extended API methods with varying semantic relations.

\subsubsection{Evaluation Datasets}
\label{dataset}
In summary, we adopt three test datasets covering three different scenarios, i.e., high-quality dataset with single ground-truth API (i.e., BIKER’s test dataset), real-wold random dataset with single ground-truth API, and 60 manual SO dataset with multiple ground-truth API.
\begin{itemize}
    \item BIKER test dataset: is the evaluation dataset of BIKER, which contains 413 manually selected and verified SO queries with API answers.
    \item Random test dataset: contains 1K random selected SO queries with API answers from BIKER’s training dataset.
    \item Multi-API SO test dataset: contains 60 manually selected SO queries with multiple relevant APIs.
\end{itemize}

\subsection{Baselines}
\label{baseline}
We compared KAHAID with BIKER \cite{Huang2018APIMR}, CLEAR \cite{Wei2022CLEARCL}, ZaCQ \cite{Eberhart2022GeneratingCQ}, which are three state-of-the-art API recommendation techniques.

Baseline1 (BIKER) \cite{Huang2018APIMR}: is an API search approach that uses a word-embedding model to calculate the semantic similarity between a query and each API description, returning the top-N API methods with similarity scores. 
Additionally, BIKER calculates the similarity score between the given query and other queries on Stack Overflow, enabling it to return similar queries and extend the given query with other related queries. 
By doing so, BIKER is capable of returning a broad range of API methods.
Its source code can be downloaded on Github\footnote{\href{https://github.com/tkdsheep/BIKER-ASE2018}{\textcolor{blue}{https://github.com/tkdsheep/BIKER-ASE2018}}}. 

Baseline2 (CLEAR) \cite{Wei2022CLEARCL}: is a API search approach based on BERT sentence embedding and contrastive learning.
CLEAR selects a set of candidate Stack Overflow posts based on BERT sentence embedding similarity and reranks them using a BERT-based classification model to recommend the top-N APIs.
Two different models were trained in CLEAR for recommending class-level APIs and method-level APIs.
Our evaluation in this work focuses solely on the performance of the API recommendation at the method-level model. 
Therefore, we solely utilize the method-level model for our experiments.
Its data and source code can be downloaded on\footnote{\href{https://github.com/Moshiii/CLEAR-replication}{\textcolor{blue}{https://github.com/Moshiii/CLEAR-replication}}}.

Baseline3 (ZaCQ) \cite{Eberhart2022GeneratingCQ}: is a source code search method that re-ranks code snippets by interactively refining queries with multiple rounds of clarifying questions.
Following ref \cite{Eberhart2022GeneratingCQ}, we select ZaCQ's Top-3 code snippets after the first round of question clarification, in which we collect all API methods as the result API methods.
Its data and source code can be downloaded on Github\footnote{\href{https://github.com/Zeberhart/ZaCQ}{\textcolor{blue}{https://github.com/Zeberhart/ZaCQ}}}.

\subsection{Performance Measures}
\label{metric}
Following existing studies \cite{Wei2022CLEARCL}, we use Mean reciprocal rank (MRR)~\cite{Radev2002EvaluatingWQ}, Mean average precision (MAP) \cite{Sanderson2010ChristopherDM}, Precision, and Recall to evaluate the performance of API recommendation approaches. 
MRR and MAP are the widely accepted measurements for information retrieval. 
MRR measures the effort needed to find the first correct answer in the recommended list and MAP considers the ranks of all correct answers.
We also evaluate the performance with Precision and Recall, which can be defined as follows:

\begin{equation}
    Precision=\frac{recommended API \cap ground-truth API}{recommended API}
\end{equation}

\begin{equation}
    Recall=\frac{recommended API \cap ground-truth API}{ground-truth API}
\end{equation}

\section{RESULT ANALYSIS}
\subsection{RQ1: How Performance is KAHAID compare with the Existing API Search Approaches~\cite{Huang2018APIMR, Wei2022CLEARCL}?}

\subsubsection{Method}
\label{rq1:method}
We compared KAHAID with two existing API search baselines, BIKER \cite{Huang2018APIMR} and CLEAR \cite{Wei2022CLEARCL}, on three different test datasets (see section~\ref{dataset}). As the authors of BIKER and CLEAR have provided replication packages, we used these to conduct experiments and compare the results.

KAHAID generates CQs with options for each dialogue round, which are utilized to recommend different API sequences depending on the selected options.
To simulate user behavior, we device an option selection strategy based on the assumption that the user always select the option that leads to the best API for their query.
This strategy selects an option only if the resulting API sequence includes the ground-truth's best API; otherwise, KAHAID's recommended API sequence is considered empty.
To conduct a fair and objective comparison between KAHAID and the baselines, we assess their top-5 API in recommended API sequence for each query.
We consider a API is correct if it is the best API in the ground-truth APIs.

\subsubsection{Result}

\vspace{-2mm}
\begin{table}[h]
\centering
\caption{Performance Comparison of API Search Approach.}
\vspace{-2mm}
\resizebox{\columnwidth}{!}{
\begin{tabular}{|c|c|c|c|c|}
\hline
Dataset                                                                              & Metric    & BIKER & CLEAR          & KAHAID          \\ \hline
\multirow{4}{*}{BIKER test dataset}                                                  & MRR       & 0.614 & \textbf{0.755} & 0.610           \\ \cline{2-5} 
                                                                                     & MAP       & 0.143 & \textbf{0.765} & 0.503           \\ \cline{2-5} 
                                                                                     & Precision & 0.249 & 0.550          & \textbf{0.693}  \\ \cline{2-5} 
                                                                                     & Recall    & 0.714 & 0.764          & \textbf{0.853}  \\ \hline
\multirow{4}{*}{Random test dataset}                                                 & MRR       & 0.253 & 0.413         & 0.428            \\ \cline{2-5} 
                                                                                     & MAP       & 0.096 & 0.366          & 0.365           \\ \cline{2-5} 
                                                                                     & Precision & 0.110 & 0.293          & \textbf{0.639}  \\ \cline{2-5} 
                                                                                     & Recall    & 0.301 & 0.397          & \textbf{0.516}  \\ \hline
\multirow{4}{*}{\begin{tabular}[c]{@{}c@{}}Multi-API SO\\ test dataset\end{tabular}} & MRR       & 0.359 & 0.523         & \textbf{0.769}  \\ \cline{2-5} 
                                                                                     & MAP       & 0.219 & 0.243          & \textbf{0.794}  \\ \cline{2-5} 
                                                                                     & Precision & 0.152 & 0.353          & \textbf{0.839}  \\ \cline{2-5} 
                                                                                     & Recall    & 0.733 & 0.251          & \textbf{0.867}  \\ \hline
\end{tabular}
}
\label{RQ:1}
\vspace{-2mm}
\end{table}

Table~\ref{RQ:1} shows the result of KAHAID compared with the other baselines.
For dataset with single ground-truth API (BIKER test dataset and Random test dataset),  overall KAHAID outperforms both BIKER and CLEAR.
Especially on the BIKER test data, KAHAID achieved a high recall of 0.853, which is an improvement of 19.5\% and 11.6\% over BIKER and CLEAR, respectively. 
This indicates that KAHAID can accurately recommend the most appropriate API for 85.3\% of query. 
Note that, Both BIKER and CLEAR have the same performance reported in this work and CLEAR`s paper.
However, different from the comparison reported in BIKER’s paper, the performance of BIKER and CLEAR was worse, especially on the Random test data.
The four metrics (MRR, MAP, Precision, Recall) of BIKER are all below 0.301, and the four metrics of CLEAR are all below 0.413.
This is because we used the same random method to get dataset as the CLEAR paper, but it may not coincide with the random test dataset in CLEAR`s paper.
Additionally, BIKER and CLEAR focus on accepted answers, which are not always the best APIs.
On the contrary, KAHAID guides users to clarify their intentions through dialogue and accurately recommend APIs that match their intentions, not just limited to accepted answers.
For example, given the question ``Getting the name of t he currently executing method \footnote{\href{https://stackoverflow.com/questions/442747/}{https://stackoverflow.com/questions/442747/}}'', KAHAID recommends the method \textit{java.lang.Class.getEnclosingMethod()} as the best API, which is present in the comment section of the highest-scoring answer rather than the accepted answer.
Instead, both BIKER and CLEAR recommend the accepted answer, i.e., \textit{java.lang.Thread.getStackTrace()}, which has a flaw as it may occasionally return a zero-length array, as described in the API documentation\footnote{\href{https://docs.oracle.com/en/java/javase/11/docs/api/java.base/java/lang/Thread.html\#getStackTrace()}{https://docs.oracle.com/en/java/javase/11/docs/api/java.base-/java/lang/Thread.html\#getStackTrace()}}.

For the Muti-API SO test data, KAHAID outperforms BIKER (by 114.2\%, 262.6\%, 452\%, and 18.3\% improvement) and CLEAR (by 47\%, 226.7\%, 137.7\%, and 245.4\% improvement) on all four evaluation metrics (MRR, MAP, Precision, Recall).
This indicates that compared to existing API search approaches, KAHAID has strong knowledge expansion ability. 
Meanwhile, we noticed that BIKER achieves a recall of 0.733, 192\% higher than CLEAR and only 15.5\% lower than KAHAID. This is because BIKER's method of calculating similarity between SO posts and APIs is more tolerant to variations in questions compared to CLEAR.
However, we do not consider this as a form of knowledge expansion, as the API sequence recommended by BIKER contains many APIs that are not related to the query, which is also the reason for its low precision.

\noindent\fbox{
\begin{minipage}{8.5cm}{
\textit{In comparison to existing API search methods BIKER and CLEAR, the human-AI dialogue method KAHAID has a strong ability for API recommendation and knowledge extension.}
}
\end{minipage}}

\vspace{-2mm}
\subsection{\textbf{RQ2}: How Performance is KAHAID compare with Existing Human-AI Dialogue Approach~\cite{Wei2022CLEARCL}}
\subsubsection{Method}
Real-world queries often have multiple correct answers that are semantically related, such as through Function Similarity or Function Collaboration. 
To address this, human-AI dialogue approaches recommend different APIs from various aspects, while also using knowledge expansion to identify APIs with semantic relations. 
To compare KAHAID with the existing human-AI dialogue approach, ZaCQ, we utilize the Multi-API SO test dataset described in section~\ref{query and APIs}. 
This dataset includes both the best API and other APIs with semantic relations to the best API.
We employ the same option selection strategy as mentioned in section~\ref{rq1:method} to simulate user behavior for both KAHAID and ZaCQ. 
Then, we calculate metrics (MRR, MAP, Precision, Recall) for the API sequences recommended by KAHAID and ZaCQ from the first round of dialogue to the third.
This enables us to evaluate and compare the performance of both human-AI dialogue approaches in recommending semantically related APIs.

\vspace{-2mm}
\subsubsection{Result}

\vspace{-2mm}
\begin{table}[h]
\centering
\caption{Performance Comparison of query clarification Approach.}
\vspace{-2mm}
\begin{tabular}{|c|c|c|c|c|}
\hline
Metrics                    & Approaches & round 1        & round 2        & round 3        \\ \hline
\multirow{2}{*}{MRR}       & KAHAID     & 0.506          & \textbf{0.769} & \textbf{0.815} \\ \cline{2-5} 
                           & ZaCQ       & \textbf{0.524} & 0.542          & 0.574          \\ \hline
\multirow{2}{*}{MAP}       & KAHAID     & 0.508          & \textbf{0.794} & \textbf{0.845} \\ \cline{2-5} 
                           & ZaCQ       & \textbf{0.521} & 0.576          & 0.582          \\ \hline
\multirow{2}{*}{Precision} & KAHAID     & 0.654          & \textbf{0.839} & \textbf{0.842} \\ \cline{2-5} 
                           & ZaCQ       & 0.500          & 0.615          & 0.615          \\ \hline
\multirow{2}{*}{Recall}    & KAHAID     & \textbf{0.721} & \textbf{0.867} & \textbf{0.875} \\ \cline{2-5} 
                           & ZaCQ       & 0.633          & 0.628          & 0.638          \\ \hline
\end{tabular}
\label{rq2}
\vspace{-2mm}
\end{table}

Table 1 shows the result of KAHAID compared with ZaCQ.
KAHAID demonstrated a significant improvement from the first to the second round of dialogue, with increases of 52\%, 56.3\%, 28.2\%, and 21.8\% in MRR, MAP, Precision, and Recall, respectively. 
The third round showed the highest values with MRR, MAP, Precision, and Recall of 0.815, 0.845, 0.842, and 0.875, which ourperfors the ZaCQ by 42\%, 45.2\%, 36.9\%, and 37.1\%.
This demonstrates that KAHAID has a strong capability for knowledge extension and can meet the diverse needs of developers.
To our surprise, ZaCQ has a poor knowledge extension, as indicated by its low MRR and MAP scores, which did not exceed 0.582. Additionally, its Precision and Recall scores were no higher than 0.638.
This is because ZaCQ is merely a tool for re-ranking initial search results through multiple rounds of clarifying conversations.
If the best API is not found in the initial search results, ZaCQ fails to obtain the best API no matter how many rounds of re-ranking are performed.
KAHAID, on the other hand, can obtain the best API via semantic relation extension even if it is not found in the initial search results.
For example, Given a query ``How do I remove repeated elements from ArrayList?'', ZaCQ cannot recommend the best API ``\textit{java.util.stream.Stream.distinct}'', either in the initial results or in any subsequent rounds of re-ranked results.
Given the same query, KAHAID could not find the best API in the initial results either.
However, by extending the API ``\textit{java.util.stream.Stream.collect}'' in the initial results, the best API can be found in the extended APIs.

\vspace{1mm}
\noindent\fbox{
\begin{minipage}{8.5cm}{
\textit{In comparison to existing human-AI dialogue baseline ZaCQ, KAHAID has a strong ability for knowledge extension; it works well even when faced with an uncertain and under-specified API query.}
}
\end{minipage}}

\vspace{-3mm}
\subsection{RQ3: How Semantically Diverse are the Question Options Generated during the Human-AI Dialogue?}
A variety of options can provide users with additional references and inspiration.
In this section, we measure the diversity of CQ options in each human-AI dialogue round.

\vspace{-3mm}
\subsubsection{Method}
For 60 experimental queries in the test set, we use KAHAID to generate multiple rounds of human-AI dialogue for each query.
Each round of dialogue includes a clarification question and question options. 
We select every option and allow KAHAID to generate different clarification questions and question options for the next round of dialogue depending on the option selected.
In this way, we can explore all possible selection path.

We invite two participants (two master students who are unaffiliated with this work but are familiar with Java) to rate the semantic diversity between two options from 0 to 1.
Score 0 indicates that there is no diversity and that the semantics of the two options are extremely similar.
Score 0.5 indicates that the two options have moderate diversity and roughly similar semantics, for example, the semantics of option B is just one of many semantics of option A.
Score 1 denotes a high degree of diversity, indicating that the semantics of the two options are completely different.

There are four types of options available: Event, Object, Object Constraint, and Event Constraint.
The types of options for the same clarification question are the same.
We score the Object, Object Constraint, or Event Constraint options according to the semantic diversity scoring criteria described above.
We don`t judge diversity solely on lexical differences, as some options have different words but have same semantics (e.g., ``path'' and ``path object'').
That is also why we manually assess diversity.

The Event option consists of verbs and objects.
The diversity score between two such options is rated based on two conditions: the verbs in options are diverse, and the two objects in options are diverse.
If two conditions are met, the diversity score is 1; if neither condition is met, the score is 0; and if only one condition is met, the score is 0.5.
We score semantic diversity of objects in the same way that we score Object options.
To score the diversity of verbs, we refer to the 87 verb classes summarized by Xie et al~\cite{Xie2020APIMR} who group verbs that have the same function into a single class.
Two verbs are diverse if they do not belong to the same class in function.


We calculate the semantic diversity of a related option in a clarification question by averaging the Semantic diversity of each related option and the other options in the clarification question. Then, for each clarification question in a dialogue round, we average the semantic diversity of question options to obtain the Semantic diversity of question options in this dialogue round. To obtain the diversity of question options in the entire query dialogue, we average the diversity of question options across multiple rounds of human-AI dialogue. Finally, we compute the overall diversity of question options in human-AI dialogue by averaging the diversity of question options across 60 query dialogues.

\vspace{-2mm}
\subsubsection{Result}
\begin{figure}[t]
    \centering
    \includegraphics[width=0.5\textwidth]{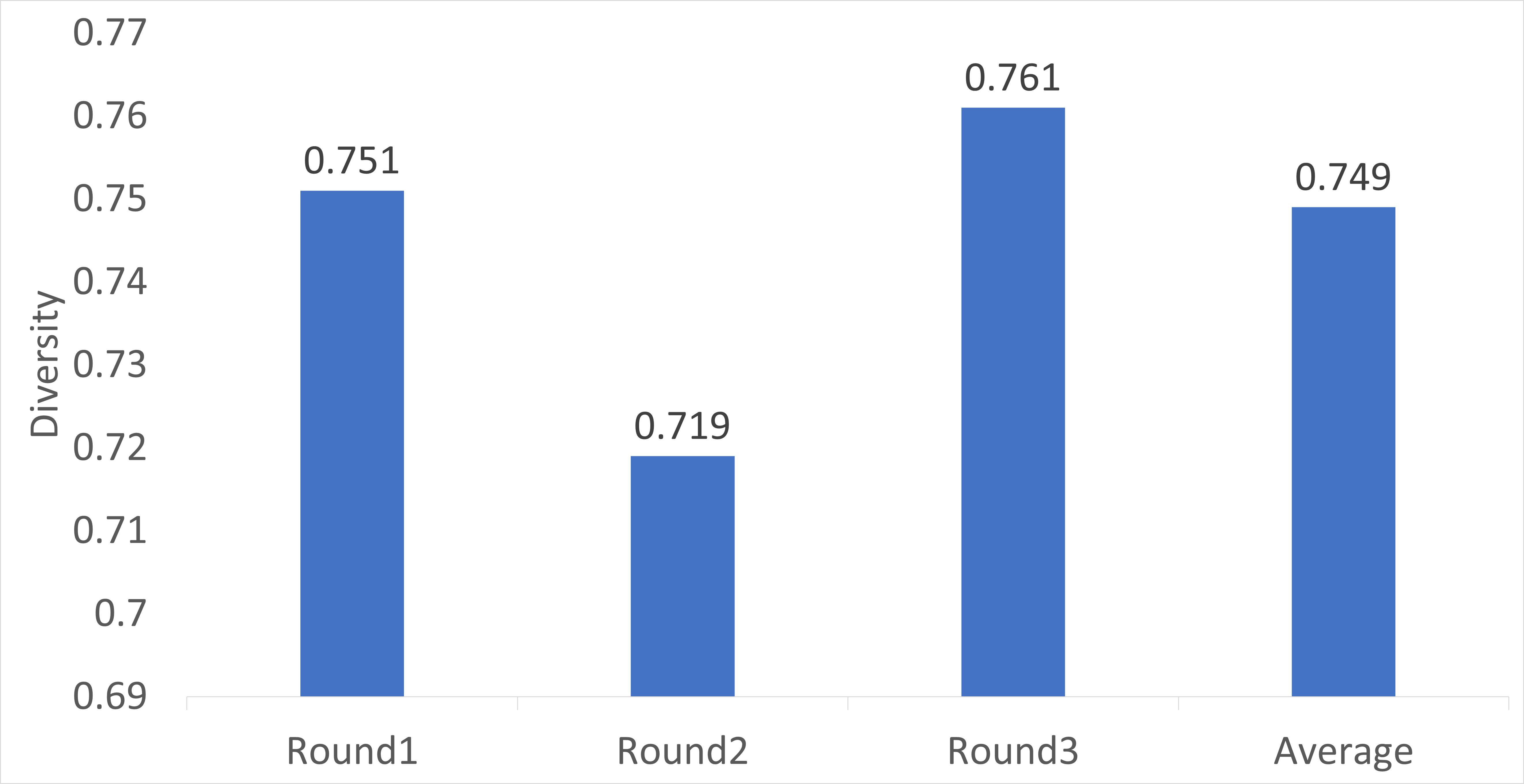}
    \caption{Diversity of Question Options in Each Round}
    \label{Fig:rq3}
    \vspace{-4mm}
\end{figure}
Fig.~\ref{Fig:rq3} depicts, around the 60 questions, the diversity of question options generated in each round of human-AI dialogue (such as 0.751 in the first round, 0.719 in the second round, and  0.761 in the third round), as well as the overall diversity of question options for the 60 questions (0.749) .
Note that not every human-AI dialogue includes three dialogue rounds; some have fewer than three rounds, while others have more than three rounds.
It is only counted to the rounds that the dialogue should have for less than three rounds, and it does not participate in the average calculation in the subsequent round. When there are more than three rounds, we only count up to the third round.

Diverse question options may encourage divergent thinking and help developers clarify what they really want..
As shown in Fig.~\ref{Fig:motivating example}, in the second CQ round, two different options (path string, path object) are generated, causing the user to realize that the initial question ``How to get the current working directory?'' is very vague and that he should rewrite the initial question more specifically into ``How to get the path object of current working directory?'' using the knowledge ``path object'' obtained from the options.

\noindent\fbox{
\begin{minipage}{8.5cm}{
\textit{During the dialogue process, KAHAID can generate a wide variety of options for each clarification question, encouraging divergent thinking and assisting developers in clarifying what he truly desires.}
} 
\end{minipage}}

\vspace{-4mm}
\subsection{RQ4: What Factors can Improve Human-AI Dialogue Efficiency?}
Human-AI dialogue efficiency is determined by the number of its rounds, which is affected by both external and internal factors.

\vspace{-3mm}
\subsubsection{Method}
The uncertainty of the API query is the external factor.
The more uncertain the query, the more rounds of dialogue are required to gradually clarify the query.
To confirm this, we create three types of uncertain and under-specified queries.
These queries are made up of four syntactic roles (verb, direct object, preposition, and preposition object) that are derived from the 25,020 API descriptions mentioned in the KG construction process.

These three types of queries are listed as follows:

(1) 18,683 ``V-DO'' queries, each with a verb (V) and a direct object (DO).
For example, ``obtain (V) data time (DO).''
(2) 6,543 ``V-PO'' queries, each with a verb (V), a preposition and a preposition object (PO).
For example, ``obtain (V) in chronology (PO).''
(3) 6,132 ``V-DO-PO'' queries, each with a verb, a direct object, a preposition and a preposition object. 
For example, ``obtain (V) date time (DO) in chronology (PO).''
This type of queries are less uncertain than the ``V-DO'' queries and ``V-PO'' queries.

In addition, we also adopt the real-world queries, i.e., 60 queries collect from SO as depict in section~\ref{query and APIs}.


The decision tree strategy is the internal factor, because the underlying decision tree determines each round of questions and options, as well as what the next round of questions ask.
We employ these two decision tree strategies (ID3 and C4.5) to support human-AI dialogue around those three types of queries and compute the average rounds of human-AI dialogue (HAR) by Eq.~\ref{eq:cps}.

\vspace{-1mm}
\begin{equation}
    HAR=\frac{\sum_{i=0}^{n} \left | Leaf_{i} \right |\times Depth_{i}}{\left | API \right | } 
    \label{eq:cps}
\end{equation}
\vspace{-1mm}
where \textit{$\left|Leaf_{i}\right|$} refers to the number of APIs contained in the i-th leaf node of this decision tree; \textit{$Depth_i$} refers to the depth of the i-th leaf node; and $\left|API\right|$ refers to the number of APIs contained in all leaf nodes of a decision tree.
In the decision tree, because the dialogue ends when receiving any API from the i-th leaf node, the dialogue clarification questions and options are generated based on non-leaf nodes (and their edges) from the root node to the leaf node.
As a result, the depth of a leaf node \textit{$Depth_i$} in the decision tree can be used to represent the number of dialogue rounds required to obtain any API in the leaf node.
Specifically, we use $ \sum_{i=0}^{n} \left | Leaf_{i} \right |\times Depth_{i} $ to compute the sum of the number of human-AI dialogue rounds required to obtain each API in a decision tree, and then make it divide the number of all APIs in the decision tree to calculate HAR.
The fewer rounds of dialogue there are on average, the more efficient the human-AI dialogue.

\subsubsection{Result}

\begin{figure}[t]
    \centering
    \includegraphics[width=0.5\textwidth]{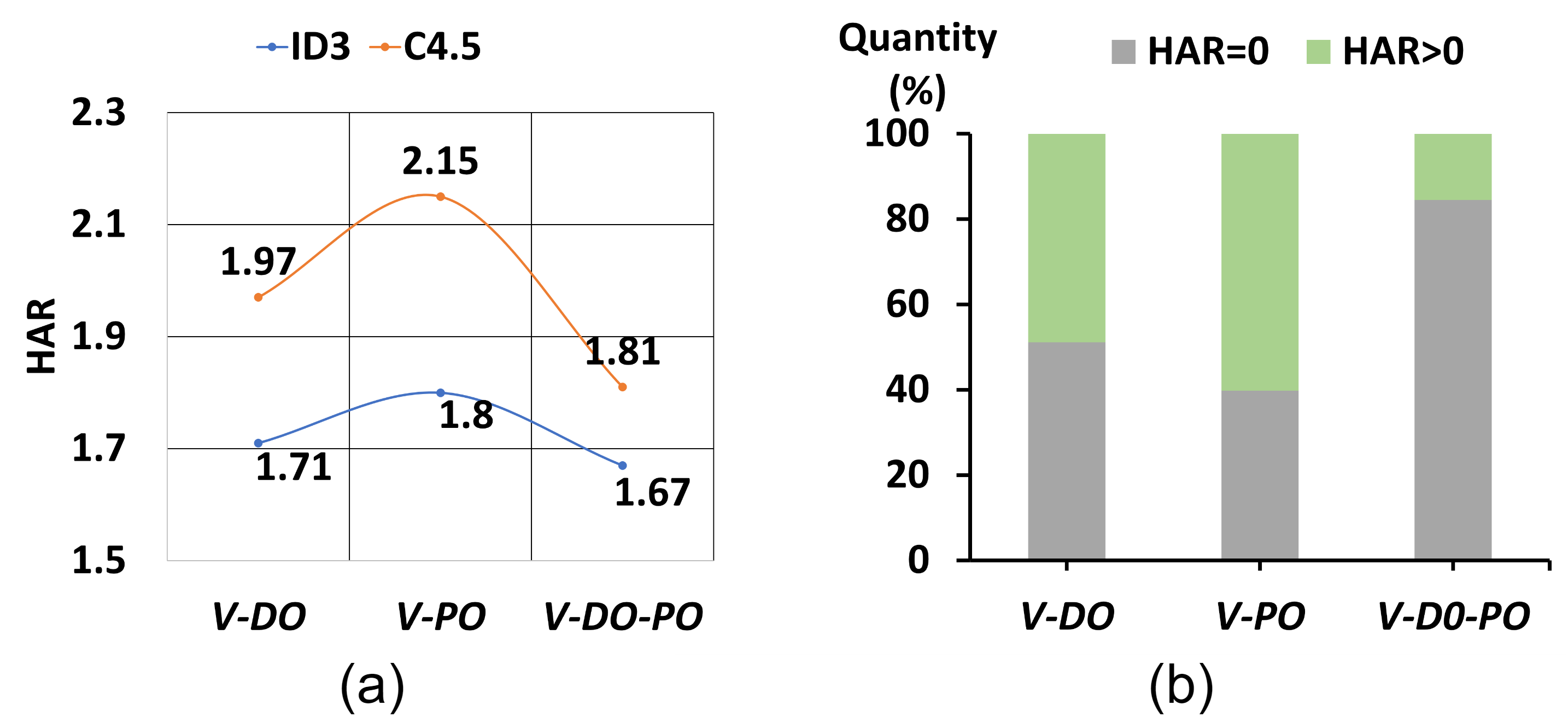}
    \caption{Internal and External Factors Influencing the Human-AI dialogue Efficiency}
    \label{Fig:rq4}
    \vspace{-2mm}
\end{figure}

As shown in Fig.~\ref{Fig:rq4}-a, regardless of the query, ID3's HAR is always lower than C4.5's and is no more than 3, which suggests that ID3 is more efficient than C4.5.
This is because C4.5 calculates an information gain ratio based on ID3 and then selects the aspect dialogue round. 
This type of CQ has fewer options, which reduces the difficulty of making decisions per dialogue round. 
However, it increases the average number of dialogue rounds, which is not conducive to guiding users to quickly clarify their intentions.
As a result, we do not use C4.5.
Furthermore, when compared with ``V-DO-PO'' queries, ``V-DO'' queries and ``V-PO'' queries have greater uncertainty, resulting in a much larger proportion of HAR greater than 0, as shown in Fig.~\ref{Fig:rq4}-b.
This shows that the more uncertain the API query, the more dialogue rounds are required.
It also implies that our method can assist users in clarifying questions through multiple dialogue rounds.

\noindent\fbox{
\begin{minipage}{8.5cm}{
\textit{The ID3 decision tree strategy improves the efficiency of question clarification and reduces the number of human-AI dialogue rounds.}
}
\end{minipage}}

\vspace{-3mm}
\section{USER STUDY}
\begin{table*}[h]
\centering
\caption{Eight Real Queries for User Study}
\resizebox{\textwidth}{!}{
\begin{tabular}{|l|l|l|l|}
\hline
PID    & StackOverflow ID & Query                                                       & Best API                                                \\ \hline
$Q_1$  & 153724           & How to round a number to n decimal places in Java?          & java.text.DecimalFormat.format                          \\ \hline
$Q_2$  & 4871051          & How to get the current working directory in Java?           & java.io.File.getAbsolutePath                            \\ \hline
$Q_3$  & 5868369          & How can I read a large text file line by line using Java?   & java.io.BufferedReader.readLine                         \\ \hline
$Q_4$  & 2860943          & How can I hash a password in Java?                          & javax.crypto.SecretKeyFactory.generateSecret            \\ \hline
$Q_5$  & 1069066          & Convert Date to String?                                     & java.lang.Thread.getStackTrace                          \\ \hline
$Q_6$  & 428918           & How can I increment a date by one day in Java?              & java.time.LocalDate.plusDays                            \\ \hline
$Q_7$  & 35842            & How can a Java program get its own process ID?              & java.lang.management.ManagementFactory.getRuntimeMXBean \\ \hline
$Q_8$  & 9481865          & Getting the IP address of the current machine using Java?   & java.net.InetAddress.getLocalHost                       \\ \hline
\end{tabular}
}
\label{tab: queries}
\end{table*}

\vspace{-2mm}
\begin{figure*}[t]
    \centering
    \includegraphics[width=0.9\textwidth]{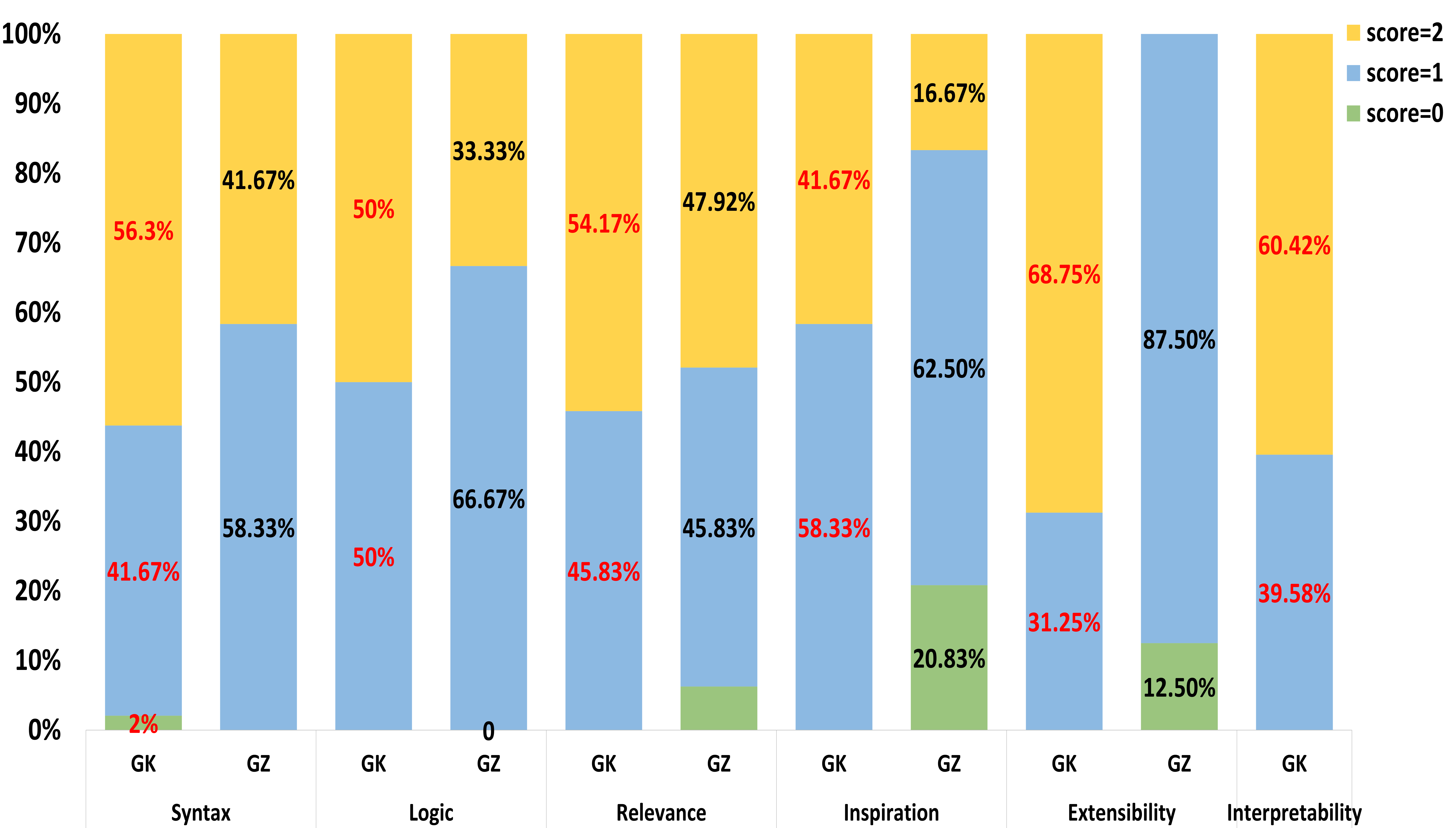}
    \footnotesize
        \begin{itemize}
            \item[] Note that the percentages of people who scored 2, 1, and 0 are represented by yellow, blue, and green, respectively.
            \item[] $G_k$ refers to the group using KAHAID; $G_z$ refers to the group using ZaCQ.
        \end{itemize}  
    \vspace{-0.3cm}
    \caption{The Proportion of Ratings Given to Each of the Six Indicators.}
    \label{Fig:user study}
    \vspace{-3mm}
\end{figure*}

\subsection{Study Design}

\subsubsection{Test Set}
8 questions and their ground-truth API methods are chosen at random from the test set described in Section \ref{query and APIs} for the user study, as shown in Table \ref{tab: queries}.

\vspace{-2mm}
\subsubsection{Participants}
We recruited 12 participants from both university and IT companies. Six of them (6 graduate students) are from the first author’s university, and six of them are from two IT companies. All of them have Java developing experience in either commercial or open source projects, and the years of their developing experience vary from 1 year to 5 years, with an average of 2.9 years.


Through a pre-study survey, we ensure that none of these students had encountered the experimental questions before.
We divided the 12 participants into two groups, with each group containing three graduate students and three participants from IT companies.
$Group{Z}$ used ZaCQ and $Group{K}$ used KAHAID.
Each group member was asked to answer 8 questions using the specific tool.

\vspace{-2mm}
\subsubsection{Study Setup}
We gave all group members a 20-minute training session to teach them how to use the specific tool because they were expected to use it to answer 8 questions.

Given the goal of this user study is to investigate the user experience with the tool rather than whether the user can use the tool to obtain the ground-truth API method, we set up the following scenario:
If the result API is the ground-truth API method, the tool tells the user that he succeeded and allows the user to answer the next question; otherwise, the tool encourages the user to interact with the tool again until the ground-truth API method is found.
If the user cannot find the ground-truth API method within 5 minutes, he will be informed that he has failed and will be permitted to answer the next question.

While answering 8 questions with the tool, we ask each group member to rate the process of answering each question in terms of the following 6 indicators:

\vspace{-2mm}
\begin{itemize}
    \item \textit{Syntax}, which measures the syntactic correctness of the generated clarification questions.
    \item \textit{Logic}, which assesses the logical correctness of the generated clarification questions.
    \item \textit{Relevance}, which measures the relationship between the generated clarification questions and the query.
    \item \textit{Inspiration}, which assesses how well the clarification question options generated enlighten the user.
    \item \textit{Extensibility}, which measures how well the extended API meets the query.
    \item \textit{Interpretability}, which measures how well the resulting API is understood in relation to each other and to the query based on API descriptions and highlighted keywords in these descriptions.Note that only $Group{K}$ are required to score this indicator because only KAHAID can explain why the resulting API was found.
\end{itemize}

\vspace{-2mm}
Each of these indicators has a score between 0 and 2, with 0 indicating dissatisfaction, 1 indicating satisfaction, and 2 indicating extreme satisfaction.
Finally, we ask each group member to write down their feedback, which include both advantages and disadvantages.

\subsection{Participants' Feedback}

Both positive and negative feedback about KAHAID are posted, respectively.

Advantages:
\vspace{-2mm}
\begin{itemize}
    \item The majority of clarification questions were syntactically and logically correct, as well as closely related to queries.
    They were extremely helpful in clarifying my query requirements.
    \item I particularly enjoyed the extended APIs because they could teach me new things. 
    The API with the efficiency comparison, in particular, surprised me by responding to the query more quickly.
    \item Various options presented during the QA process deeply inspired me to solve problems in previously unknown ways.
    \item These API descriptions and highlighted API keywords were useful. 
    They explained how APIs related to queries and what APIs could do, giving me more confidence in my final decision.
\end{itemize}

\vspace{-2mm}
Disadvantages:
\vspace{-2mm}
\begin{itemize}
    \item Some clarifying questions, while grammatically and logically correct, are not expressed naturally.
    This problem adds some comprehension time to the questions.
    \item The presentation of result APIs is a little perplexing. The layout is clumsy, especially when showing multiple result APIs at once, which makes my experience unpleasant.
\end{itemize}

\vspace{-1.5mm}
\subsection{Indicator Analysis}
We collected each group member's rating results on different indicator, and plotted them in Fig~\ref{Fig:user study}.
This figure shows that $G_k$ outperforms $G_Z$ across the indicators. 
Although more than 90\% of $G_k$ and $G_Z$ gave Syntax, Logic, and Relevance scores greater than 0, the percentage of $G_k$ with a score of 2 was higher than $G_Z$ (56.3\% vs. 41.67\%, 54.17\% vs. 47.92\%, 50\% vs. 33.33\%).
In terms of Inspiration and Extensibility, the $G_k$ and $G_Z$ score results are significantly different.
Compared to ZaCQ, on scores of 1 and 2, KAHAID improves by 20.83\% Inspiration and by 12.5\% Extensibility.
Through analyzing the options participants selected during human-AI dialogue process as well as the result API they obtained, we also have the following findings:

First, participants prefer options that inspire them to think of new ways to answer the questions.
For example, $G_k$ assigned a higher Inspiration score to Q2.
This illustrates that the options generated by KAHAID are more illuminating than those generated by ZaCQ.
Specifically, ZaCQ generated a clarification question for Q2: ``Are you interested in getting the current working directory for the Default File System?'' with only two options ``Yes'' or ``No.''
As a result, $G_z$ were not inspired beyond the ``default file system'' strategy.
In contrast, $G_k$ were inspired by the diverse options generated by KAHAID in the first round (``return filesystem'', ``return path'' and ``converting path string to path Object'').
$G_k$ discovered that, in addition to ``return filesystem'', they could get the current working directory by ``returning path'' and ``converting path string to path Object''.
KAHAID, in particular, employs the ``path object'' option to encourage $G_k$ to obtain the current working path via the Path Object.
Finally, 83.33\% of $G k$ participants chose this option and assigned a score of 2 to Q2.

Second, participants prefer the extended APIs that remind them of implicit knowledge they were previously unaware of.
The Extensibility score for these extended APIs was frequently higher. 
For Q5, for example, KAHAID recommended an extended API ``\textit{java.text.DateFormat.format}'', which has a Efficiency Comparison relation with the best API ``\textit{java.lang.String.format}''. 
This extended API provides $G_k$ with a faster response to the query, which most participants were previously unaware of. 
As a result, 75\% of $G_k$ gave Q5 a Extensibility score of 2 and 25\% gave it a score of 1.
Although the APIs suggested by ZaCQ can also be used to answer this query, they are all well-known to $G_Z$ participants.
Finally, all of the $G z$ gave Q5 Extensibility scores of 1.

Third,  if the API method is explainable, participants can find the best API method more easily or with greater confidence.
Q7, for example, received a 100\% Interpretability score of 2 from $G_k$.
Both KAHAID and ZaCQ can find its ground-truth API method  ``\textit{ManagementFactory.getRuntimeMXBean}'' for Q7.
However, due to the API's lack of interpretability, half of $G_z$ could not associate it with the query simply by its name.
In contrast, $G_k$ can has access to the API method interpretation, which includes the API descriptions and highlighted API keywords.
Following that, $G_k$ understands how this API method relates to the query through the keyword (``return managed bean'' and  ``runtime system''), and what it is capable of according to the API description.
Eventually, $G_k$ found this API method and was convinced that it was the best API method.

\vspace{1mm}
\noindent\fbox{
\begin{minipage}{8.5cm}{
\textit{KAHAID can provide a good user experience based on user ratings and feedback in six areas: syntax, logic, relevance, inspiration, extensibility, and interpretability.}
} 
\end{minipage}}

\vspace{-3.0mm}
\section{THREATS TO VALIDITY}
One \textbf{intrinsic threat} is the use of the AllenNLP, a general-purpose natural language processing library.
AllenNLP is used to parse sentences in both the semantic and syntactic analyses of the API function descriptions.
However, none of the tools, including AllenNLP, can perfectly parse large amounts of textual data.
Furthermore, AllenNLP is not specifically designed to parse API function descriptions containing code elements, incomplete sentences, or common syntax errors.
To mitigate this threat, we devise heuristics for adapting AllenNLP to parsing API function descriptions.
AllenNLP, for example, has trouble parsing API descriptions that frequently begin with a verb.
As a result, we devise a heuristic rule: such incomplete sentences are prefixed with ``This method'', which aids AllenNLP in parsing the sentence components.
For another example, the content in parentheses is useless, which affects AllenNLP's ability to parse the sentence.
As a result, we device a heuristic rule that automatically removes the content in parentheses in the sentence, improving AllenNLP's ability to parse the sentence.


One \textbf{external threat} is the generality of our API behavior KG.
Because we built it entirely from official Java API documentation, the KG cannot play a role in a broader domain, such as the domain of other languages.
In the future, we plan to expand KG with additional programming language documentation, allowing it to handle a broader range of programming issues.

\vspace{-2mm}
\section{RELATED WORK}
API search refers to the process of obtaining APIs related to a search query using various modeling techniques \cite{Liu2022OpportunitiesAC} such as information retrieval models \cite{Wu2008InterpretingTT}, machine learning models, and deep learning models \cite{Huang2020ACR,Huang2018APIMR, Cai2021SearchFC}.
If the search query is too broad, these search models are unlikely to find the relevant APIs.
To make the query more specific, many query expansion techniques have been proposed, such as expanding query words with relevant software artifacts from official API documents \cite{Lv2015CodeHowEC, Hu2020UnsupervisedSR}, code changes \cite{Huang2019EnhanceCS, Huang2019QEintegratingFB, Huang2019DeepLT, Huang2018QueryEB}, or Stack Overflow posts \cite{Hu2020UnsupervisedSR, Nie2016QueryEB, Sirres2018AugmentingAS, Zhang2018ExpandingQF}.
While this enables a more specific query and the discovery of the most relevant API, it lacks interpretation and extensibility, making it difficult for developers to interpret search results and preventing them from discovering potentially useful APIs.
Furthermore, current API search research is still a long way from social-technical information seeking on online forums \cite{Zhang2021ReadingAO}, where pragmatic API needs suggestive, explainable, and extensible API recommendation, such as exploring alternative or better solutions, and discovering previously unknown API knowledge \cite{Ren2020DemystifyOA, Eberhart2021DialogueMF}.

To meet practical API needs, we propose KAHAID with intent clarification, result interpretation, and extension capabilities.
It differs from current API search research in three ways.
1) In terms of enlightenment, KAHAID interacts with the developer to clarify actions, objects, or constraints until it finds some APIs.
This may motivate him to seek a better solution or to determine his true desires and rewrite the initial vague question, which is not supported by current query expansion methods.
Even though ZaCQ \cite{Eberhart2022GeneratingCQ} is a conversational search method, the various aspects of the clarification questions it generates are limited to only the syntactic analysis for the tasks associated with the search results.
In contrast, the diverse options for the clarification questions generated by KAHAID are derived from the API behavior knowledge graph, which contains rich API actions, objects, constraints, and various API semantic relations. 
Furthermore, ZaCQ is merely a tool for re-ranking the initial search results. If the initial search results do not contain the correct API, no matter how many QAs the developer runs, there is no way to find the correct API.
Under such conditions, KAHAID, on the other hand, can use KG to extend the correct API by following the relationships between APIs.
2) In terms of interpretation, KAHAID displays the discovered APIs along with a concise explanation of why these APIs are recommended and how they relate to the clarified query.  Specifically, KAHAID displays the API's functionality description and highlights the keywords it uses for clarification, as well as the relations between the extend APIs and the most relevant API.
3) In terms of extensibility, KAHAID provides a variety of APIs, including directly relevant APIs and extended APIs, based on the knowledge graph's rich API semantic relations.

\vspace{-2mm}
\section{Conclusion and Future Work}
In this paper, we propose KAHAID as an initial attempt to combine the immediate responsiveness of API search technologies with the interaction, clarification, explanation, and extensibility capability of social-technical information seeking.
The systematic evaluation confirms that KAHAID implements an illuminating, interpretable, and extensible exploratory query process.

In the future, in addition to Java SDK APIs, we will implement KAHAID for more APIs, such as Python APIs, Ruby APIs and Go APIs.
Furthermore, if KAHAID can teach a user to search an API via Human-KG dialog, we may be able to teach the LLM (Large pre-trained Language Model) \cite{Ouyang2022TrainingLM,Solaiman2021ProcessFA} to adapt the specific downstream task via LLM-KG dialog.
$ \langle ab$This will make KGQA (Knowledge Graph-based Question Answering) more open and flexible.

\section*{Acknowledgements}
The work is partly supported by the National Natural Science Foundation of China under Grant (Nos. 61902162, 62262031), the Natural Science Foundation of Jiangxi Province (20202BAB202015), and the Graduate Innovative Special Fund Projects of Jiangxi Province(YC2021-S308, YC2022-s258).

\normalem
\balance
\bibliography{sample-base}

\begin{thebibliography}{10}
\providecommand{\url}[1]{#1}
\csname url@samestyle\endcsname
\providecommand{\newblock}{\relax}
\providecommand{\bibinfo}[2]{#2}
\providecommand{\BIBentrySTDinterwordspacing}{\spaceskip=0pt\relax}
\providecommand{\BIBentryALTinterwordstretchfactor}{4}
\providecommand{\BIBentryALTinterwordspacing}{\spaceskip=\fontdimen2\font plus
\BIBentryALTinterwordstretchfactor\fontdimen3\font minus
  \fontdimen4\font\relax}
\providecommand{\BIBforeignlanguage}[2]{{%
\expandafter\ifx\csname l@#1\endcsname\relax
\typeout{** WARNING: IEEEtran.bst: No hyphenation pattern has been}%
\typeout{** loaded for the language `#1'. Using the pattern for}%
\typeout{** the default language instead.}%
\else
\language=\csname l@#1\endcsname
\fi
#2}}
\providecommand{\BIBdecl}{\relax}
\BIBdecl

\bibitem{Ren2020DemystifyOA}
X.~Ren, J.~Sun, Z.~Xing, X.~Xia, and J.~Sun, ``Demystify official api usage
  directives with crowdsourced api misuse scenarios, erroneous code examples
  and patches,'' \emph{2020 IEEE/ACM 42nd International Conference on Software
  Engineering (ICSE)}, pp. 925--936, 2020.

\bibitem{Eberhart2021DialogueMF}
Z.~Eberhart and C.~McMillan, ``Dialogue management for interactive api
  search,'' \emph{2021 IEEE International Conference on Software Maintenance
  and Evolution (ICSME)}, pp. 274--285, 2021.

\bibitem{Haiduc2011OnTE}
S.~Haiduc and A.~Marcus, ``On the effect of the query in ir-based concept
  location,'' \emph{2011 IEEE 19th International Conference on Program
  Comprehension}, pp. 234--237, 2011.

\bibitem{Thung2013AutomaticRO}
F.~Thung, S.~Wang, D.~Lo, and J.~L. Lawall, ``Automatic recommendation of api
  methods from feature requests,'' \emph{2013 28th IEEE/ACM International
  Conference on Automated Software Engineering (ASE)}, pp. 290--300, 2013.

\bibitem{Rahman2016RACKAA}
M.~M. Rahman, C.~K. Roy, and D.~Lo, ``Rack: Automatic api recommendation using
  crowdsourced knowledge,'' \emph{2016 IEEE 23rd International Conference on
  Software Analysis, Evolution, and Reengineering (SANER)}, vol.~1, pp.
  349--359, 2016.

\bibitem{Mikolov2013DistributedRO}
T.~Mikolov, I.~Sutskever, K.~Chen, G.~S. Corrado, and J.~Dean, ``Distributed
  representations of words and phrases and their compositionality,'' in
  \emph{NIPS}, 2013.

\bibitem{Ye2016FromWE}
X.~Ye, H.~Shen, X.~Ma, R.~C. Bunescu, and C.~Liu, ``From word embeddings to
  document similarities for improved information retrieval in software
  engineering,'' \emph{2016 IEEE/ACM 38th International Conference on Software
  Engineering (ICSE)}, pp. 404--415, 2016.

\bibitem{Huang2018APIMR}
Q.~Huang, X.~Xia, Z.~Xing, D.~Lo, and X.~Wang, ``Api method recommendation
  without worrying about the task-api knowledge gap,'' \emph{2018 33rd IEEE/ACM
  International Conference on Automated Software Engineering (ASE)}, pp.
  293--304, 2018.

\bibitem{Gu2016DeepAL}
X.~Gu, H.~Zhang, D.~Zhang, and S.~Kim, ``Deep api learning,'' \emph{Proceedings
  of the 2016 24th ACM SIGSOFT International Symposium on Foundations of
  Software Engineering}, 2016.

\bibitem{Xie2020APIMR}
W.~Xie, X.~Peng, M.~Liu, C.~Treude, Z.~Xing, X.~Zhang, and W.~Zhao, ``Api
  method recommendation via explicit matching of functionality verb phrases,''
  \emph{Proceedings of the 28th ACM Joint Meeting on European Software
  Engineering Conference and Symposium on the Foundations of Software
  Engineering}, 2020.

\bibitem{Zhang2021ReadingAO}
H.~Zhang, S.~Wang, T.-H.~P. Chen, and A.~E. Hassan, ``Reading answers on stack
  overflow: Not enough!'' \emph{IEEE Transactions on Software Engineering},
  vol.~47, pp. 2520--2533, 2021.

\bibitem{Ren2019DiscoveringEA}
X.~Ren, Z.~Xing, X.~Xia, G.~Li, and J.~Sun, ``Discovering, explaining and
  summarizing controversial discussions in community q\&a sites,'' \emph{2019
  34th IEEE/ACM International Conference on Automated Software Engineering
  (ASE)}, pp. 151--162, 2019.

\bibitem{Liu2021APIRelatedDI}
M.~Liu, X.~Peng, A.~Marcus, S.~Xing, C.~Treude, and C.~Zhao, ``Api-related
  developer information needs in stack overflow,'' \emph{IEEE Transactions on
  Software Engineering}, 2021.

\bibitem{Liu2020GeneratingCB}
Y.~Liu, M.~Liu, X.~Peng, C.~Treude, Z.~Xing, and X.~Zhang, ``Generating concept
  based api element comparison using a knowledge graph,'' \emph{2020 35th
  IEEE/ACM International Conference on Automated Software Engineering (ASE)},
  pp. 834--845, 2020.

\bibitem{Bonial2010PropBankAG}
C.~Bonial, O.~Babko-Malaya, J.~D. Choi, and J.~D. Hwang, ``Propbank annotation
  guidelines,'' 2010.

\bibitem{Huang2022112PK}
Q.~Huang, Z.~Yuan, Z.~Xing, Z.~Zuo, C.~Wang, and X.~Xia, ``1+1>2: Programming
  know-what and know-how knowledge fusion, semantic enrichment and coherent
  application,'' \emph{ArXiv}, vol. abs/2207.05560, 2022.

\bibitem{Gardner2018AllenNLPAD}
M.~Gardner, J.~Grus, M.~Neumann, O.~Tafjord, P.~Dasigi, N.~F. Liu, M.~E.
  Peters, M.~Schmitz, and L.~Zettlemoyer, ``Allennlp: A deep semantic natural
  language processing platform,'' \emph{ArXiv}, vol. abs/1803.07640, 2018.

\bibitem{Eberhart2022GeneratingCQ}
Z.~Eberhart and C.~McMillan, ``Generating clarifying questions for query
  refinement in source code search,'' in \emph{SANER}, 2022.

\bibitem{CastleGreen2020DecisionTA}
T.~Castle-Green, S.~Reeves, J.~E. Fischer, and B.~Koleva, ``Decision trees as
  sociotechnical objects in chatbot design,'' \emph{Proceedings of the 2nd
  Conference on Conversational User Interfaces}, 2020.

\bibitem{Hssina2014ACS}
B.~Hssina, A.~Merbouha, H.~Ezzikouri, and M.~Erritali, ``A comparative study of
  decision tree id3 and c4.5,'' \emph{International Journal of Advanced
  Computer Science and Applications}, vol.~4, 2014.

\bibitem{Quinlan2004InductionOD}
J.~R. Quinlan, ``Induction of decision trees,'' \emph{Machine Learning},
  vol.~1, pp. 81--106, 2004.

\bibitem{Quinlan1992C45PF}
{J. Ross Quinlan}, ``C4.5: Programs for machine learning,'' 1992.

\bibitem{Wei2022CLEARCL}
M.~Wei, N.~S. Harzevili, Y.~Huang, J.~Wang, and S.~Wang, ``Clear: Contrastive
  learning for api recommendation,'' \emph{2022 IEEE/ACM 44th International
  Conference on Software Engineering (ICSE)}, pp. 376--387, 2022.

\bibitem{Radev2002EvaluatingWQ}
D.~R. Radev, H.~Qi, H.~Wu, and W.~Fan, ``Evaluating web-based question
  answering systems,'' in \emph{International Conference on Language Resources
  and Evaluation}, 2002.

\bibitem{Sanderson2010ChristopherDM}
M.~Sanderson, ``Christopher d. manning, prabhakar raghavan, hinrich
  sch{\"u}tze, introduction to information retrieval, cambridge university
  press 2008. isbn-13 978-0-521-86571-5, xxi + 482 pages,'' \emph{Natural
  Language Engineering}, vol.~16, pp. 100 -- 103, 2010.

\bibitem{Liu2022OpportunitiesAC}
C.~Liu, X.~Xia, D.~Lo, C.~Gao, X.~Yang, and J.~C. Grundy, ``Opportunities and
  challenges in code search tools,'' \emph{ACM Computing Surveys (CSUR)},
  vol.~54, pp. 1 -- 40, 2022.

\bibitem{Wu2008InterpretingTT}
H.~C. Wu, R.~W.~P. Luk, K.-F. Wong, and K.-L. Kwok, ``Interpreting tf-idf term
  weights as making relevance decisions,'' \emph{ACM Trans. Inf. Syst.},
  vol.~26, pp. 13:1--13:37, 2008.

\bibitem{Huang2020ACR}
Q.~Huang, A.~Qiu, M.~Zhong, and Y.~Wang, ``A code-description representation
  learning model based on attention,'' \emph{2020 IEEE 27th International
  Conference on Software Analysis, Evolution and Reengineering (SANER)}, pp.
  447--455, 2020.

\bibitem{Cai2021SearchFC}
F.~Cai, C.~Wang, Q.~Huang, Z.~Zuo, and Y.~Liao, ``Search for compatible source
  code,'' \emph{Int. J. Softw. Eng. Knowl. Eng.}, vol.~31, pp. 477--502, 2021.

\bibitem{Lv2015CodeHowEC}
F.~Lv, H.~Zhang, J.-G. Lou, S.~Wang, D.~Zhang, and J.~Zhao, ``Codehow:
  Effective code search based on api understanding and extended boolean model
  (e),'' \emph{2015 30th IEEE/ACM International Conference on Automated
  Software Engineering (ASE)}, pp. 260--270, 2015.

\bibitem{Hu2020UnsupervisedSR}
G.~Hu, M.~Peng, Y.~Zhang, Q.~Xie, W.~GAO, and M.~Yuan, ``Unsupervised software
  repositories mining and its application to code search,'' \emph{Software:
  Practice and Experience}, vol.~50, pp. 299 -- 322, 2020.

\bibitem{Huang2019EnhanceCS}
Q.~Huang and G.~Wu, ``Enhance code search via reformulating queries with
  evolving contexts,'' \emph{Automated Software Engineering}, vol.~26, pp. 705
  -- 732, 2019.

\bibitem{Huang2019QEintegratingFB}
Q.~Huang and H.~Wu, ``Qe-integrating framework based on github knowledge and
  svm ranking,'' \emph{Science China Information Sciences}, vol.~62, pp. 1--16,
  2019.

\bibitem{Huang2019DeepLT}
Q.~Huang, Y.~Yang, and M.~Cheng, ``Deep learning the semantics of change
  sequences for query expansion,'' \emph{Software: Practice and Experience},
  vol.~49, pp. 1600 -- 1617, 2019.

\bibitem{Huang2018QueryEB}
Q.~Huang, Y.~Yang, X.~Zhan, H.~Wan, and G.~Wu, ``Query expansion based on
  statistical learning from code changes,'' \emph{Software: Practice and
  Experience}, vol.~48, pp. 1333 -- 1351, 2018.

\bibitem{Nie2016QueryEB}
L.~Nie, H.~Jiang, Z.~Ren, Z.~Sun, and X.~Li, ``Query expansion based on crowd
  knowledge for code search,'' \emph{IEEE Transactions on Services Computing},
  vol.~9, pp. 771--783, 2016.

\bibitem{Sirres2018AugmentingAS}
R.~Sirres, T.~F. Bissyand{\'e}, D.~Kim, D.~Lo, J.~Klein, K.~Kim, and Y.~L.
  Traon, ``Augmenting and structuring user queries to support efficient
  free-form code search,'' \emph{Empirical Software Engineering}, vol.~23, pp.
  2622--2654, 2018.

\bibitem{Zhang2018ExpandingQF}
F.~Zhang, H.~Niu, I.~Keivanloo, and Y.~Zou, ``Expanding queries for code search
  using semantically related api class-names,'' \emph{IEEE Transactions on
  Software Engineering}, vol.~44, pp. 1070--1082, 2018.

\bibitem{Ouyang2022TrainingLM}
L.~Ouyang, J.~Wu, X.~Jiang, D.~Almeida, C.~L. Wainwright, P.~Mishkin, C.~Zhang,
  S.~Agarwal, K.~Slama, A.~Ray, J.~Schulman, J.~Hilton, F.~Kelton, L.~E.
  Miller, M.~Simens, A.~Askell, P.~Welinder, P.~F. Christiano, J.~Leike, and
  R.~J. Lowe, ``Training language models to follow instructions with human
  feedback,'' \emph{ArXiv}, vol. abs/2203.02155, 2022.

\bibitem{Solaiman2021ProcessFA}
I.~Solaiman and C.~Dennison, ``Process for adapting language models to society
  (palms) with values-targeted datasets,'' in \emph{NeurIPS}, 2021.

\end{thebibliography}

\vspace{-3mm}
\par\noindent 
\parbox[t]{\linewidth}{
\noindent\parpic{\includegraphics[height=3.0in,width=1in,clip,keepaspectratio]{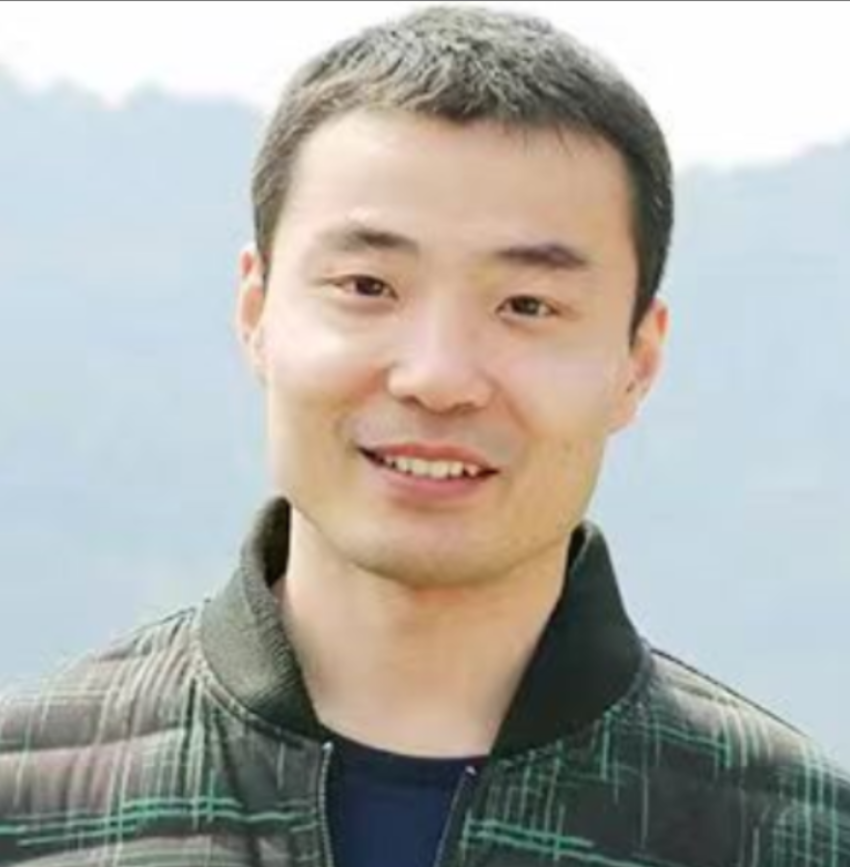}}
\noindent {\bf QING HUANG} 
received the M.S degree in computer application and technology from Nanchang University, in 2009, and the PH.D. degree in computer software and theory from Wuhan University, in 2018. He is currently an Assistant Professor with the School of Computer and Information Engineering, Jiangxi Normal University, China. His research interests include information security, software engineering and knowledge graph.}

\par\noindent 
\parbox[t]{\linewidth}{
\noindent\parpic{\includegraphics[height=3.0in,width=1in,clip,keepaspectratio]{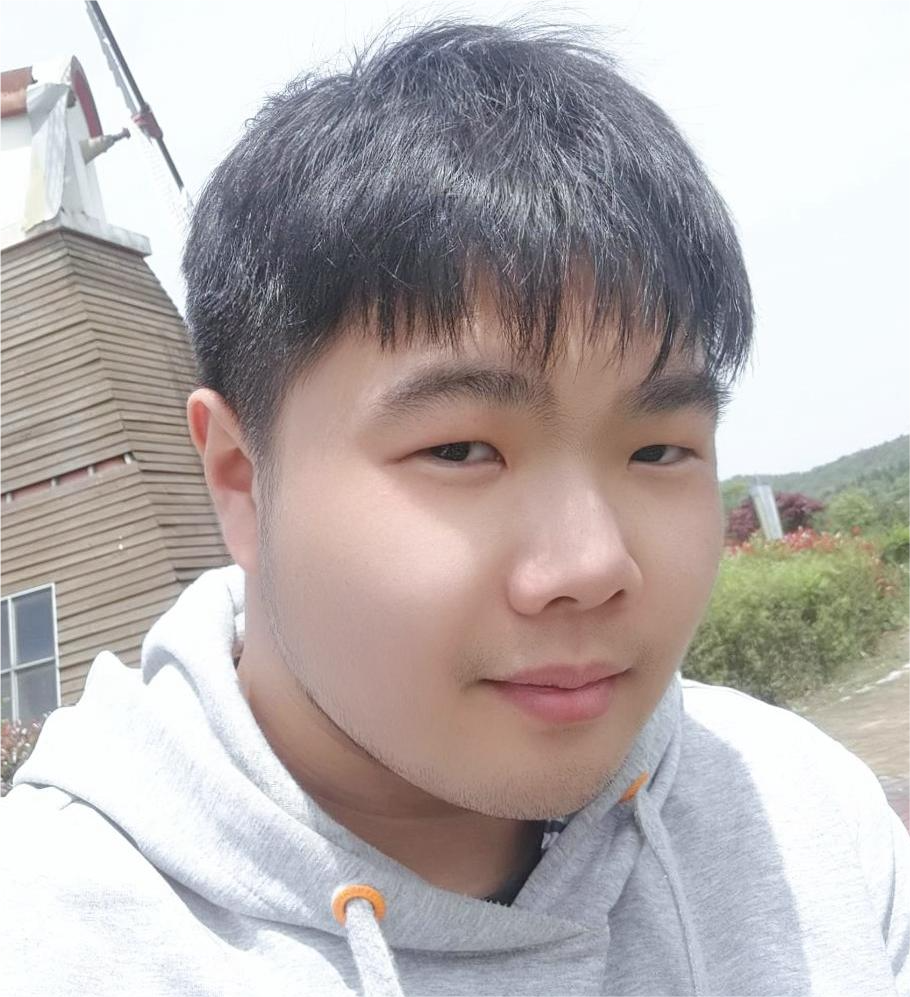}}
\noindent {\bf Zishuai Li}\ 
is a second-year graduate student in the School of Computer and Information Engineering, Jiangxi Normal University, China. His research interests are software engineering, human-computer interaction and knowledge graph.
}
\vspace{0.5\baselineskip}

\par\noindent 
\parbox[t]{\linewidth}{
\noindent\parpic{\includegraphics[height=3.0in,width=1in,clip,keepaspectratio]{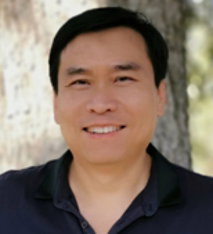}}
\noindent {\bf Zhenchang Xing}\
is a Senior Research Scientist with Data61, CSIRO, Eveleigh, NSW, Australia. In addition, he is an Associate Professor in the Research School of Computer Science, Australian National University. Previously, he was an Assistant Professor in the School of Computer Science and Engineering, Nanyang Technological University, Singapore, from 2012-2016. His main research areas are software engineering, applied data analytics, and human-computer interaction.}

\par\noindent 
\parbox[t]{\linewidth}{
\noindent\parpic{\includegraphics[height=3.0in,width=1in,clip,keepaspectratio]{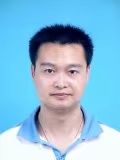}}
\noindent {\bf ZHENGKANG ZUO}\
received the Ph.D. degree in computer science and technology from the Chinese Academy of Sciences (CAS), Beijing, China. He is currently a Associate Professor and deputy director of the Computer Science and Technology Department of Jiangxi Normal University, Nanchang, China. His main research interests include software formal methods, generic programming, etc.}

\par\noindent 
\parbox[t]{\linewidth}{
\noindent\parpic{\includegraphics[height=3.0in,width=1in,clip,keepaspectratio]{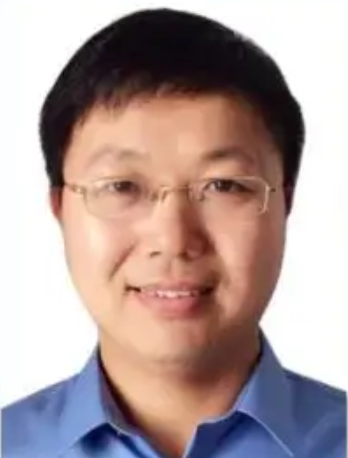}}
\noindent {\bf Xin Peng}\
is a professor with the School of Computer Science, Fudan University, China. His research interests include data-driven intelligent software development, cloud-native software and AIOps, and software engineering for AI and cyber-physical-social systems. He was the recipient of the ICSM 2011 Best Paper Award, the ACM SIGSOFT Distinguished Paper Award at ASE 2018, the IEEE TCSE Distinguished Paper Awards at ICSME 2018,2019,2020, and IEEE Transactions on Software Engineering 2018 Best Paper Award.}

\par\noindent 
\parbox[t]{\linewidth}{
\noindent\parpic{\includegraphics[height=3.0in,width=1in,clip,keepaspectratio]{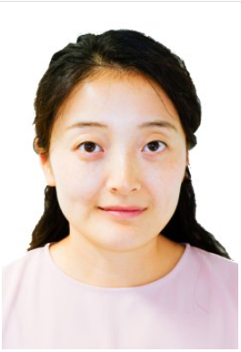}}
\noindent {\bf Xiwei Xu}\
is a Senior Research Scientist with Architecture\& Analytics Platforms Team, Data61, CSIRO. She is also a Conjoint Lecturer with UNSW. She started working on blockchain since 2015. She is doing research on blockchain from software architecture perspective, for example, tradeoff analysis, and decision making and evaluation framework. Her main research interest is software architecture. She also does research in the areas of service computing, business process, and cloud computing and dependability.}

\par\noindent 
\parbox[t]{\linewidth}{
\noindent\parpic{\includegraphics[height=3.0in,width=1in,clip,keepaspectratio]{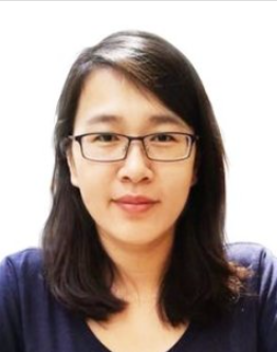}}
\noindent {\bf Qinghua Lu}\
is a Senior Research Scientist with Data61, CSIRO, Eveleigh, NSW, Australia. Before joining Data61, she was an Associate professor at China University of Petroleum, and she worked as a researcher at National Information and Communications Technology Australia. She has published more than 100 academic papers in international journals and conferences. Her research interests include the software architecture, blockchain, software engineering for AI, and AI ethics.}

\end{document}